\documentstyle[12pt,epic,eepic]{article}
\def\tr{\mbox{tr}}
\def\Sp{\mbox{Sp}}

\begin{document}
\begin{flushright}
\begin{tabular}{l}
KOBE-TH-97-02\hspace{0.5cm}\\
September 1997
\end{tabular}
\end{flushright}
\vspace*{5mm}
\begin{center}
{\large \bf Dynamical symmetry breaking in the external gravitational 
and constant magnetic fields}\\[6mm]
T.~Inagaki
\footnote{e-mail : inagaki@hetsun1.phys.kobe-u.ac.jp},\\
{\it Department of Physics, Kobe University, 
Rokkoudai, Nada, Kobe 657, Japan}\\[3mm]
S.~D.~Odintsov
\footnote{e-mail : sergei@ecm.ub.es, odintsov@quantum.univalle.edu.co},\\
{\it Dept. Theor. Phys., Tomsk Pedagogical University,
634041 Tomsk, Russia\\
and\\
Department de Fisica, Universidad del Valle,
A.A.25360 Cali, Colombia}\\[3mm]
Yu. I. Shil'nov
\footnote{e-mail : visit2@ieec.fcr.es}\\
{\it Institut D'Estudis Espacials De Catalunya,\\
Edif. Nexus-104, Gran Capita 2-4, 08034, Barcelona, Spain\\
and\\
Department of Theoretical Physics, Faculty of Physics,\\
Kharkov State University,
Svobody Sq. 4, 310077, Kharkov, Ukraine}\\[8mm]
\end{center}

\begin{abstract}
We investigate the effects of the external gravitational 
and constant magnetic fields to the dynamical symmetry
breaking. As simple models of the dynamical symmetry
breaking we consider the Nambu-Jona-Lasinio (NJL) model 
and the supersymmetric Nambu-Jona-Lasinio (SUSY NJL) model 
non-minimally interacting with the external gravitational
field and minimally interacting with constant magnetic field.
The explicit expressions for the scalar and spinor Green 
functions are found up to the linear terms on the spacetime 
curvature and exactly for  a constant magnetic field.
We obtain the effective potential of the above models from 
the Green functions in the magnetic field in curved spacetime. 
Calculating the effective potential numerically with the 
varying curvature and/or magnetic fields we show the effects 
of the external gravitational and magnetic fields to the phase 
structure of the theories. In particular, increase of the curvature
in the  spontaneously broken chiral symmetry phase due to the fixed
magnetic field makes this phase to be less broken. 
On the same time strong magnetic field quickly induces chiral 
symmetry breaking even at the presence of fixed gravitational field
within nonbroken phase. 
\end{abstract}

\newpage

\section{Introduction}

The idea of the dynamical symmetry breaking in quantum field theory
has been introduced quite long ago \cite{NJL}.
In the attempts to realize this idea in the early universe
one should study the external gravitational field (or curved
spacetime). The investigation of the dynamical symmetry breaking 
in four-fermion models \cite{NJL,GN} in four and three dimensional
curved spacetime has been started in Refs.\cite{IMO,EOS1} 
using the weak curvature approximation
(for a review
and list of references, see \cite{rev}). 
A validity of the weak curvature approximation is discussed
in Ref.\cite{IN}.
In the case when the 
external gravitational field is treated exactly (de Sitter
or anti-de Sitter or Einstein universe background) 
the phase structure of the 
four-fermion models has been studied in Refs.\cite{ELOS}.
(for a renormalization group approach, see also \cite{HS}).

There appeared recently some indications that early universe may
contain large primordial magnetic fields.
The role of these fields in cosmology (in particularly, in the 
inflationary models of the universe) has been discussed 
in Refs.\cite{M1}.
The presence of magnetic field in the early universe may
lead to different effects.
In particular, it is reasonable then to investigate the dynamical
symmetry breaking in curved spacetime with magnetic fields%
\footnote{For the  works where four-fermion models have been
discussed in the presence of constant electromagnetic fields,
see \cite{M2,K1,K2}}.
Such study for 3D and 4D four-fermion models
has been undertaken in Refs.\cite{GOS,GGO} in the approximation when
the effects of the gravitational and magnetic field maybe simply
summarized (in other words, the correspondent terms in the effective
potential are evaluated with the neglecting of the interaction
between gravitational and magnetic fields).
However, the coherent effect of combined gravitational and magnetic
field may become very relevant.

That is the purpose of the present work to investigate the phase 
structure of four-fermion models in weakly curved spacetime with 
constant magnetic field. We develop the approximation where the
effective potential and Green function (GF) of the theory
maybe represented as the expansion in the powers of curvature 
invariants. In each order of this expansion the external magnetic
field is treated exactly, i.e. we take into account the coherent
effect of combined gravitational and magnetic fields.

The paper is organized as follows.
In the next section we present new form of the local momentum
representation of propagators \cite{BP} with account of the
constant magnetic field. In each order on the expansion in curvature 
invariants for Green functions the magnetic field is included exactly.
The detailed calculation is presented for scalar and spinor Green 
functions. Section 3 is devoted to the study of phase structure in
the NJL model in curved spacetime with magnetic field.
We calculate the effective potential with account of linear curvature
terms for 3D and 4D models.
The numerical investigation of the effective potential for 4D NJL
model is given. In section 4 we discuss the SUSY
NJL model non-minimally interacting with the external gravitational
field and minimally interacting with constant magnetic field.
The effective potential is evaluated and the correspondent phase
structure is given. Some discussion and outlook are presented in
the last section.

\section{Scalar and spinor Green functions in the magnetic field
in curved spacetime}

One of the fundamental objects in the field theory
is a Green function. It is a basic quantity for a number of 
calculations of quantum effects.
Here we calculate the scalar and spinor
Green functions in the magnetic field in curved spacetime
to study the phase structure of the NJL and SUSY NJL
model in the external gravitational and constant magnetic
fields.

\subsection{Green function for a scalar field}

First we study  GF for the 
scalar field $G(x,y;m)$ in an external electromagnetic 
and gravitational field. We assume that the spacetime
curves slowly and neglect the terms involving the
metric derivatives higher than third
(weak curvature approximation). However we treat the
external magnetic field exactly. The GF 
for the scalar field $G(x,y;m)$ satisfies the Klein-Gordon 
equation :
\begin{equation}
     (iD^{\mu}iD_{\mu}-m^{2}-\xi R)G(x,y;m)
     =\frac{1}{\sqrt{-g}}\delta^{D}(x-y),
\label{Gfunc}
\end{equation}
where $\xi$ represents the non-minimal coupling constant 
with external gravitational field(see \cite{BOS}), 
$m$  is the scalar field mass  
and $D^{\mu}=\nabla^{\mu}-ieA^{\mu}$.
In the present paper the vector potential of the external 
electromagnetic field is chosen in the form 
\begin{equation}
A_\mu(x)=-{1 \over 2 } F_{\mu\nu} x^\nu ,
\label{def:A}
\end{equation}
where $F_{\mu\nu}$ is constant matrix of electromagnetic field
strength tensor.
The covariant derivative for a scalar field 
is only an ordinary derivative $\nabla^{\mu}=\partial^{\mu}$.
For a vector field $v^{\mu}$ the covariant derivative
is rewritten as
\begin{eqnarray}
     \nabla_{\mu}v^{\mu}&=&\partial_{\mu}v^{\mu}
     +{\Gamma^{\nu}}_{\nu\mu}v^{\mu}
     \nonumber \\
     &=& \frac{1}{\sqrt{-g}}\partial_{\mu}\sqrt{-g}v^{\mu}.
\end{eqnarray}
Thus the first term of the left hand side in Eq.(\ref{Gfunc}) is
rewritten as
\begin{eqnarray}
     \sqrt{-g}D^{\mu}D_{\mu}G(x,y;m)&=&
     \sqrt{-g}(\nabla^{\mu}-ieA^{\mu})
     (\nabla_{\mu}-ieA_{\mu})G(x,y;m) \nonumber \\
     &=& (\partial_{\mu}-ieA_{\mu})\sqrt{-g}g^{\mu\nu}
         (\partial_{\nu}-ieA_{\nu})G(x,y;m) .
\label{DDG}
\end{eqnarray}
We want to expand Eq.(\ref{Gfunc}) around $R=0$.
For this purpose we introduce the Riemann normal 
coordinates \cite{Ein}. 
In the Riemann normal coordinates framework
the metric tensor is expanded as 
\begin{eqnarray}
g_{\mu\nu}(y)&=& \displaystyle
\eta_{\mu\nu}+{1\over 3 } R^{(0)}_{\mu\rho\sigma\nu}(y-x)^\rho 
(y-x)^\sigma ,\nonumber \\
g(y)&=&-1-{1\over 3 } R^{(0)}_{\mu\nu}(y-x)^\mu (y-x)^\nu , 
\label{RNCE1G}
\end{eqnarray}
where the suffix $(0)$ for $R^{(0)}_{\mu\rho\sigma\nu}$
and $R^{(0)}_{\mu\nu}$
designates the curvature tensor at the origin $x$.
Substituting Eq.(\ref{RNCE1G}) into Eq.(\ref{DDG}) we obtain
\begin{eqnarray}
     \sqrt{-g}D^{\mu}D_{\mu}G(x,y;m)&=&
     \left\{\left[\eta^{\mu\nu}
     +\frac{1}{6}{R^{(0)}}_{\alpha\beta}
     (y-x)^{\alpha}(y-x)^{\beta}\eta^{\mu\nu}
     \right.\right.
     \nonumber \\
     &&\left.
     -\frac{1}{3}{{{{R^{(0)}}^{\mu}}_{\alpha}}^{\nu}}_{\beta}
     (y-x)^{\alpha}(y-x)^{\beta}\right](\partial_{\mu}-ieA_{\mu})
     \nonumber \\
     &&\left.
     -\frac{2}{3}{{R^{(0)}}^{\nu}}_{\alpha}(y-x)^{\alpha}\right\}
     (\partial_{\nu}-ieA_{\nu})G(x,y;m),
\label{DDG:RNC}
\end{eqnarray}
where we keep only terms independent of curvature
or linear in curvature.
We expand the GF, $G(x,y;m)$, as
\begin{equation}
     G(x,y;m)=G^{(0)}(x,y;m)+G^{(1)}(x,y;m)+{\cal O}(R^{2}),
\label{exp:G}
\end{equation}
where $G^{(0)}$ and $G^{(1)}$ represent the terms independent of 
$R$ and the terms linear in $R$ respectively.

Substituting Eqs.(\ref{DDG:RNC}) and (\ref{exp:G}) 
into (\ref{Gfunc}) we can perturbatively solve
the Klein-Gordon equation (\ref{Gfunc}) about the
spacetime curvature $R$.
The piece independent of curvature in Eq.(\ref{Gfunc}) is given by
\begin{equation}
     \left[\eta^{\mu\nu}(\partial_{\mu}-ieA_{\mu})
     (\partial_{\nu}-ieA_{\nu})+m^{2}\right]G^{(0)}(x,y;m)
     =-\delta^{D}(x-y).
\label{GfuncOR0}
\end{equation}
Thus $G^{(0)}(x,y;m)$ satisfies the Klein-Gordon equation
in flat spacetime. We will give the explicit expression 
for $G^{(0)}(x,y;m)$ below.
The piece linear in $R$ in Eq.(\ref{Gfunc}) becomes
\begin{eqnarray}
     &&\left\{\left[\frac{1}{6}{R^{(0)}}_{\alpha\beta}
     (y-x)^{\alpha}(y-x)^{\beta}\eta^{\mu\nu}
     \right.\right.\nonumber \\
     &&\left.-\frac{1}{3}{{{{R^{(0)}}^{\mu}}_{\alpha}}^{\nu}}_{\beta}
     (y-x)^{\alpha}(y-x)^{\beta}\right](\partial_{\mu}-ieA_{\mu})
     (\partial_{\nu}-ieA_{\nu})\nonumber \\
     &&\left.
     -\frac{2}{3}{{R^{(0)}}^{\nu}}_{\alpha}(y-x)^{\alpha}
     (\partial_{\nu}-ieA_{\nu})+\xi R \right\}G^{(0)}(x,y;m)
     \nonumber \\
     &&+\left[\eta^{\mu\nu}(\partial_{\mu}-ieA_{\mu})
     (\partial_{\nu}-ieA_{\nu})+m^{2}\right]G^{(1)}(x,y;m)=0.
\label{GfuncOR1}
\end{eqnarray}
Hence the piece $G^{(1)}$ involving the terms linear in $R$
is expressed with the help of 
the GF $G^{(0)}(x,y;m)$ in
flat spacetime (for more details of local momentum representation,
see \cite{rev,BP,Ein,PT}).

It is more convenient to introduce the new variable 
defined by
\begin{equation}
     G(x,y;m)\equiv \Phi(x,y)\tilde{G}(x-y;m),
\label{tildeG}
\end{equation}
where $\Phi(x,y)$ satisfies
\begin{equation}
     (\partial_{\mu}-ieA_{\mu})\Phi(x,y)=0.
\label{def:phi}
\end{equation}
Inserting Eq.(\ref{tildeG}) into Eq.(\ref{GfuncOR1})
we obtain
\begin{eqnarray}
     &&\left[\frac{1}{6}{R^{(0)}}_{\alpha\beta}
     (y-x)^{\alpha}(y-x)^{\beta}\partial^{\mu}\partial_{\mu}
     \right.\nonumber \\
     &&-\frac{1}{3}{{{{R^{(0)}}^{\mu}}_{\alpha}}^{\nu}}_{\beta}
     (y-x)^{\alpha}(y-x)^{\beta}\partial_{\mu}\partial_{\nu}
     \nonumber \\
     &&\left.
     -\frac{2}{3}{{R^{(0)}}^{\nu}}_{\alpha}(y-x)^{\alpha}
     \partial_{\nu}+\xi R^{(0)} \right]\tilde{G}^{(0)}(y-x;m)
     \nonumber \\
     &&+\left(\partial_{\mu}\partial_{\mu}
     +m^{2}\right)\tilde{G}^{(1)}(x-y;m)=0.
\label{tildeGfuncOR1}
\end{eqnarray}
$A_{\mu}$-dependence disappears in the relationship
between $\tilde{G}^{(0)}(y-x;m)$ and 
$\tilde{G}^{(1)}(x-y;m)$.
Therefore the GF $\tilde{G}^{(1)}(x-y;m)$
is given by
\begin{eqnarray}
     \tilde{G}^{(1)}(x-y;m)&=&\int d^{D}z G_{00}(x-z;m)
     \nonumber \\
     &&\times \left[\frac{1}{6}{R^{(0)}}_{\alpha\beta}
     (z-y)^{\alpha}(z-y)^{\beta}\partial^{\mu}\partial_{\mu}
     \right.\nonumber \\
     &&-\frac{1}{3}{{{{R^{(0)}}^{\mu}}_{\alpha}}^{\nu}}_{\beta}
     (z-y)^{\alpha}(z-y)^{\beta}\partial_{\mu}\partial_{\nu}
     \nonumber \\
     &&\left.
     -\frac{2}{3}{{R^{(0)}}^{\nu}}_{\alpha}(z-y)^{\alpha}
     \partial_{\nu}+\xi R^{(0)} \right]\tilde{G}^{(0)}(z-y;m),
\label{tildeG1}
\end{eqnarray}
where the function $G_{00}(x-y;m)$ satisfies
\begin{equation}
     (\partial^{\mu}\partial_{\mu}+m^{2})G_{00}(x-y;m)
      =-\delta^{D}(x-y).
\label{defG00}
\end{equation}
The linear curvature correction terms of the scalar GF
are expressed by the GF in flat spacetime.

In the constant curvature spacetime,
\begin{equation}
     R_{\mu\nu\rho\sigma}=\frac{R}{D(D-1)}
     (\eta_{\mu\rho}\eta_{\nu\sigma}
     -\eta_{\mu\sigma}\eta_{\nu\rho}),
\label{Rconst}
\end{equation}
the equation (\ref{tildeG1}) simplifies to
\begin{eqnarray}
     &&\tilde{G}^{(1)}(x-y;m)\\
     &&=\frac{R}{D(D-1)}\int d^{D}z G_{00}(x-z;m)
     \nonumber \\
     &&\times \left[\frac{D-3}{6}\eta_{\alpha\beta}
     (z-y)^{\alpha}(z-y)^{\beta}\partial^{\mu}\partial_{\mu}
     \right.
     +\frac{1}{3}
     (z-y)^{\mu}(z-y)^{\nu}\partial_{\mu}\partial_{\nu}
     \nonumber \\
     &&\left.
     -\frac{2}{3}(D-1)(z-y)^{\mu}
     \partial_{\mu}+D(D-1)\xi \right]\tilde{G}^{(0)}(z-y;m).
\label{tildeG1Rconst}
\end{eqnarray}

As is shown in Eq.(\ref{GfuncOR0}), 
$G^{(0)}(x,y;m)$ satisfies the Klein-Gordon equation in flat 
spacetime. According to the Schwinger proper-time method the 
Klein-Gordon equation is exactly solvable in the constant 
electromagnetic field \cite{Sch,IZ}%
\footnote{
For other types of exact solutions of the Klein-Gordon equation
in the external electromagnetic field, see \cite{BG}.
}.
To calculate $G^{(0)}(x,y;m)$ we introduce the proper-time 
Hamiltonian which is defined by
\begin{equation}
     H(x^{\mu},i\partial_{\mu})G^{(0)}(x,y;m)\equiv
     \left[(i\partial^{\mu}+eA^{\mu})
     (i\partial_{\mu}+eA_{\mu})-m^{2}\right]G^{(0)}(x,y;m).
\label{defH}
\end{equation}
%
For this proper-time Hamiltonian
the time evolution operator $U(x,y;s)$ is defined by
\begin{equation}
     i\frac{\partial}{\partial s}U(x,y;s)
     =H(x,i\partial_{\mu})U(x,y;s),
\label{defU}
\end{equation}
with the boundary conditions
\begin{equation}
\left\{
\begin{array}{lcl}
     \lim_{s\rightarrow 0} U(x,y;s)&=&\delta^{D}(x-y),\\
     \lim_{s\rightarrow -\infty} U(x,y;s)&=&0.
\end{array}
\right.
\label{bc:Gsc}
\end{equation}
%
Comparing Eq.(\ref{defU}) with Eq.(\ref{defH}) we find
\begin{equation}
     G^{(0)}(x,y;m)=-i\int^{0}_{-\infty} ds U(x,y;s).
\label{G:comp}
\end{equation}
%
%

The time evolution operator $U(x,y;s)$ is obtained
by solving the equation of motion for $x^{\mu}$
and $\pi^{\mu}=i\partial^{\mu}-eA^{\mu}$.
For a constant electromagnetic field,
$x^{\mu}(s)=U^{\dagger}(s)x^{\mu}U(s)$ and 
$\pi^{\mu} (s)=U^{\dagger}(s)\pi^{\mu}U(s)$
satisfy the following equations :
\begin{equation}
     \frac{dx^{\mu}(s)}{ds}=i[H,x^{\mu}]
     =-2\pi^{\mu} .
\end{equation}
\begin{equation}
     \frac{d\pi_{\mu}(s)}{ds}=i[H,\pi_{\mu}]
     =2eF_{\mu\nu}\pi^{\nu} .
\end{equation}
These differential equations can be easily solved,
\begin{equation}
     \pi (s)=e^{2eFs}\pi(0) ,
\label{sol:pi}
\end{equation}
\begin{equation}
     x(s)-x(0)=-\frac{e^{2eFs}-1}{eF}\pi(0),
\label{sol:x}
\end{equation}
where we use the matrix notation.
Substituting Eqs.(\ref{sol:pi}) and (\ref{sol:x})
into Eq.(\ref{defU}) we obtain the differential
equation for the time evolution operator $U(x,y;s)$,
%
%
\begin{eqnarray}
     &&i\frac{\partial}{\partial s}U(x,y;s)
     =(\pi^{2}(s)-m^{2})U(x,y;s)
     \nonumber \\
     &&=\left\{(x-y)K(x-y)
     -\frac{i}{2}\tr[eF\coth(eFs)]
        -m^{2}\right\}U(x,y;s),
\label{defU2}
\end{eqnarray}
where $K$ is defined by
\begin{equation}
     K \equiv \frac{1}{4} e^{2}F^{2}[\sinh(eFs)]^{-2}.
\end{equation}
After the integration over $s$ in Eq.(\ref{defU2}) the time 
evolution operator is found to be :
\begin{eqnarray}
     U(x,y;s)&=&-\frac{i}{(4\pi)^{D/2}}\Phi(x,y)s^{-D/2}\exp\left\{
     -\frac{1}{2}\tr \ln \left[\frac{\sinh(eFs)}{eFs}\right]
     \right\} \nonumber \\
     && \times \exp \left[\frac{i}{4}(x-y)eF \coth (eFs)(x-y)
     +im^{2}s\right].
\label{sol:U}
\end{eqnarray}
Substituting this time evolution operator (\ref{sol:U}) 
to Eq.(\ref{G:comp}) we find the GF, $G^{(0)}(x,y;m)$, 
\begin{eqnarray}
     &&G^{(0)}(x,y;m)=-i\int^{0}_{-\infty} ds U(x,y;s)
     \nonumber \\
     &&= -\frac{1}{(4\pi)^{D/2}}\Phi(x,y)\int^{0}_{-\infty}ds s^{-D/2}
     \exp\left[
     -\frac{1}{2}\tr \ln \left(\frac{\sinh(eFs)}{eFs}\right)
     \right] \nonumber \\
     && \times \exp \left[\frac{i}{4}(x-y)eF \coth (eFs)(x-y)
     +im^{2}s\right] .
\label{sol:G}
\end{eqnarray}
%
Taking the limit $F\rightarrow 0$ and changing the
variable $s \to -s$ in Eq.(\ref{sol:G})
we get the GF, $G_{00}(x-y;m)$,  
\begin{eqnarray}
     &&G_{00}(x-y;m)
     = -\int_{0}^{\infty}ds \frac{1}{(4\pi s)^{D/2}}
     \nonumber \\
     &&\times\exp \left[-i\frac{\pi}{4}D
     -\frac{i}{4s}(x-y)^{\mu}(x-y)_{\mu}
     -im^{2}s\right] ,
\label{sol:G00}
\end{eqnarray}
where the phase factor is determined to satisfy the
boundary condition (\ref{bc:Gsc}).

In the present paper we consider the constant
magnetic field along the $z$-axis.
For this constant magnetic field, $F_{12}=-F_{21}=B$,
the GF, $\tilde{G}^{(0)}(x-y;m)$,
reduces to
\begin{eqnarray}
     &&\tilde{G}^{(0)}(x-y;m)
     =-\int_{0}^{\infty}ds \frac{1}{(4\pi s)^{D/2}}
     \nonumber \\
     && \times
     \frac{eBs}{\sin(eBs)}
     \exp \left[-i\frac{\pi}{4}D
     -\frac{i}{4s}(x-y)^{\mu}C_{\mu\nu}(x-y)^{\nu}
     -i m^{2}s\right] ,
\label{sol:tildeG0M}
\end{eqnarray}
where 
\begin{equation}
C_{\mu\nu}=\eta_{\mu\nu} + F_{\mu}{}^{\lambda} F_{\lambda\nu}
{1-eBs\cot (eBs) \over B^2}.
\label{def:Cmn}
\end{equation}

Inserting the Eqs.(\ref{sol:G00}) and (\ref{sol:tildeG0M}) 
into (\ref{tildeG1Rconst}) we obtain the GF linear 
in $R$
\begin{eqnarray}
     &&\tilde{G}^{(1)}(x-y;m)
     =\frac{R}{D(D-1)}
     \int d^{D}z \int_{0}^{\infty}\frac{dt}{(4\pi t)^{D/2}}
     \int_{0}^{\infty}\frac{ds}{(4\pi s)^{D/2}}
     \nonumber \\
     &&\times
     \exp \left[-i\frac{\pi}{4}D-\frac{i}{4t}(x-z)^{\mu}(x-z)_{\mu}
     -im^{2}t\right]
     \nonumber \\
     &&\times \left[\frac{D-3}{6}\eta_{\alpha\beta}
     (z-y)^{\alpha}(z-y)^{\beta}\partial^{\mu}\partial_{\mu}
     \right.\nonumber \\
     &&\left.+\frac{1}{3}
     (z-y)^{\mu}(z-y)^{\nu}\partial_{\mu}\partial_{\nu}
     -\frac{2}{3}(D-1)(z-y)^{\mu}
     \partial_{\mu}+D(D-1)\xi \right]
     \nonumber \\
     &&\times\frac{eBs}{\sin(eBs)}
     \exp \left[-i\frac{\pi}{4}D
     -\frac{i}{4s}(z-y)^{\mu}C_{\mu\nu}(z-y)^{\nu}
     -im^{2}s\right] .
\end{eqnarray}
Thus, we obtain the explicit expression of the scalar 
GF in the external gravitational and 
magnetic field.

We need only the coincidence limit $x \to y$
of the GF to calculate the effective potential of the models 
considered in the present paper.
After the Wick rotation $s \to -is, t \to -it,
z^{0} \to -iz^{0}$ and the integration over $z$
$\Sp G(x,x;m)$ simplifies to
\begin{equation}
     \Sp G^{(0)}(x,x;m)=
     \frac{-i}{(4\pi)^{D/2}}\int_{0}^{\infty}ds s^{-D/2}
     \frac{eBs}{\sinh(eBs)}
     \exp \left(-m^{2}s\right) ,
\label{SpG0}
\end{equation}
\begin{eqnarray}
     &&\Sp G^{(1)}(x,x;m)=
     \frac{i}{(4\pi)^{D/2}}\frac{R}{D(D-1)}
     \int_{0}^{\infty} \int_{0}^{\infty} dt\ ds 
     \exp \left[-m^{2}(s+t)\right]\nonumber \\
     &&\times\frac{(s+t)^{(2-D)/2}}
     {1+eBt\coth(eBs)}\frac{eB}{\sinh(eBs)}
     \nonumber \\
     &&\times \left\{\left[
     -\frac{D-3}{6}
     \left(D-2+2eBs\coth(eBs)\right)
     +\frac{-2D+1}{3}\right]\frac{(D-2)t}{s+t}\right.
     \nonumber \\
     &&+\left[
     -\frac{D-3}{6s}
     \left(D-2+2eBs\coth(eBs)\right)
     +\frac{-2D+1}{3}eB\coth(eBs)\right]
     \nonumber \\
     &&\times\frac{2t}{1+eBt\coth(eBs)}
     \nonumber \\
     &&+\frac{2(D-1)}{3}\left[\frac{D(D-2)}{4}
     \left(\frac{t}{s+t}\right)^{2}
     +2\left(\frac{eBt\coth(eBs)}{1+eBt\coth(eBs)}\right)^{2}\right]
     \nonumber \\
     &&+\frac{1}{3s}
     \left[(D-3)\left((eBs)^{2}\coth^{2}(eBs)+1\right)
     +4eBs\coth(eBs)\right]
     \nonumber \\
     &&\times\left. \frac{(D-2)t}{s+t}
     \frac{t}{1+eBt\coth(eBs)}+D(D-1)\xi\right\} .
\label{SpG1:FIN}
\end{eqnarray}
Eqs.(\ref{SpG0}) and (\ref{SpG1:FIN}) correspond to the
the vacuum self-energy of the free scalar field with
mass $m$ at the one loop level.

\subsection{Green function for a spinor field}

Next we construct the spinor GF
in an external electromagnetic and gravitational field.
Let us write the GF, which obeys the 
Dirac equation:
\begin{equation}
(i \gamma^\mu D_\mu-m) S(x,y;m)
=\frac{1}{\sqrt{-g}}\delta^{D}(x-y),
\label{GF:spn}
\end{equation}
where the covariant derivative $D_{\mu}$ includes the electromagnetic
potential $A_{\mu}$:
\begin{equation}
D_{\mu}=\partial_{\mu} -i e A_\mu+{1 \over 2 }\omega^{ab}{}_{\!\mu} 
\sigma_{ab}.
\label{covdel:GFspn}
\end{equation}
The local Dirac matrices $\gamma_\mu (x)$ are expressed through the usual 
flat ones $\gamma_a$ and tetrads $e^a_\mu$:
\begin{equation}
\begin{array}{lcr}
\gamma^\mu (x)&=&\gamma^a e^\mu_a (x),\\
\sigma_{ab}&=& \displaystyle{1\over 4 }[\gamma_a,\gamma_b].
\end{array}
\label{gamma:GFspn}
\end{equation}
The spin-connection has the form :
\begin{eqnarray}
\omega^{ab}{}_{\!\mu}=\frac{1}{2}e^{a 
\nu}(\partial_{\mu}e^{b}_{\nu}-\partial_{\nu}
e^{b}_{\mu})+\frac{1}{4}e^{a 
\nu}e^{b\rho}e_{c\mu}(\partial_{\rho}e^{c}_{\nu}
-\partial_{\nu} e^{a}_{\rho})\nonumber \\
-\frac{1}{2}e^{b\nu}(\partial_{\mu}e^{a}_{\nu}-\partial_{\nu} 
e^{a}_{\mu})-
\frac{1}{4}e^{b\nu}e^{a\rho}e_{c\mu}(\partial_{\rho}e^{c}_{\nu}-\partial_{\nu}
 e^{c}_{\rho}).
\label{spin:conn}
\end{eqnarray}
Dimensions of the spinor 
representation are supposed to be four. 
Greek and Latin indices correspond to the curved and flat tangent
spacetimes.

To calculate the linear curvature corrections the local momentum
expansion formalism as in the previous subsection is the most 
convenient one \cite{rev,BP,PT}. 
In the Riemann normal coordinates framework
the tetrads ${e^{\mu}}_{a}(x)$ and the spin connection 
$\omega^{ab}{}_{\!\mu}\sigma_{ab}$ are expanded as 
\begin{equation}
\begin{array}{rcl}
{e^\mu}_ a (y)&=& \displaystyle
{\delta^\mu}_a+{1\over 6 } R^{(0) \mu}{}_{\!\rho\sigma a}(y-x)^\rho 
(y-x)^\sigma,\\
\omega^{ab}{}_{\!\mu}\sigma_{ab}&=& \displaystyle
{1\over 2 }R^{(0) ab}{}_{\!\mu\lambda} 
(y-x)^\lambda\sigma_{ab} .
\end{array}
\label{RNCE2}
\end{equation}
Substituting (\ref{spin:conn}) and (\ref{RNCE2}) 
into the (\ref{covdel:GFspn}), we obtain
the following equation for the GF :
\begin{eqnarray}
&&\biggl[ i\gamma^a \left({\delta^\mu}_a
+{1\over 6}R^{(0) \mu}{}_{\!\rho\sigma a}(y-x)^\rho (y-x)^\sigma
\right)
\left(\partial_\mu+{1 \over 
4}R^{(0)}_{bc\mu\lambda}(y-x)^\lambda\sigma^{bc}-ieA_\mu\right)
\nonumber \\
&&-m\biggr] S(x,y;m)=\delta^{D}(x-y).
\label{GF2:spn}
\end{eqnarray}

Fulfilling the expansion on the spacetime curvature degrees
\begin{equation}
     S=S^{(0)}+S^{(1)}+\cdots ,
\label{GFE:spn}
\end{equation}
where  $S^{(0)}$ is the GF in the flat spacetime,
$S^{(1)} \sim {\cal O}(R)$ and so on, we receive the iterative 
sequence of equations:
\begin{equation}
\biggl[ i\not{\!\partial}+e\not{\!\!A(x)}-m\biggr] 
S^{(0)}(x,y;m)=\delta^{D}(x-y)
\label{GF0:spn}
\end{equation}
\begin{eqnarray}
&&\biggl[ (i\not{\!\partial}+e 
\not{\!\!A}(x)-m)S^{(1)}(x,y;m)
\nonumber \\
&&+{i\over 6}
\gamma^a R^{(0) \mu}{}_{\!\rho\sigma a}(y-x)^\rho 
(y-x)^\sigma (\partial_\mu-i e 
A_\mu (x))\nonumber \\
&&+{i\over 4}\gamma^a R^{(0)}_{bca\lambda} 
(y-x)^\lambda \sigma^{bc} \biggr]
S^{(0)}(x,y;m)=0 .
\label{GF1:spn}
\end{eqnarray}
Here and below we can forget about the difference between the two kinds 
of indices (Greek and
Latin) because it lies beyond of linear curvature approximation.

We assume that just as in the flat spacetime GF has the form \cite{Sch}:
\begin{equation}
S(x,y;m)=\Phi(x,y)\tilde{S}(x-y;m),
\label{Eq:G:spn}
\end{equation}
where the function $\Phi(x,y)$ is introduced in Eq.(\ref{def:phi}).
Then, we can find the equation which determines the 
$\tilde{S}^{(1)}(x-y;m)$
function, excluding the evident dependence on $A_{\mu}(x)$:
\begin{eqnarray}
&&(i\not{\!\partial}-m)\tilde{S}^{(1)}(x-y;m)
\nonumber \\
&&=-{i \over 6 
}\gamma^a
R^{(0) \mu}{}_{\!\rho\sigma a} (y-x)^{\rho} (y-x)^\sigma 
\partial_\mu \tilde{S}^{(0)}(x-y;m)\nonumber \\
&&-{i \over 4}\gamma^a \sigma^{bc} R^{(0)}_{bca\lambda}
(y-x)^\lambda 
\tilde{S}^{(0)}(x-y;m) .
\label{Eq:G0:spn}
\end{eqnarray}

The flat spacetime GF in the external electromagnetic field
is supposed to be known \cite{K1,K2,Sch}. 
Denoting as $ S_{00}(x-y;m)$  the GF, 
satisfying the equation
\begin{equation}
(i\not{\!\partial}-m)S_{00}(x-y;m)=\delta^{D}(x-y),
\label{Eqn:G00:spn}
\end{equation}
we obtain
\begin{eqnarray}
&&\int d^{D}z S_{00}^{-1}(x-z;m)\tilde{S}^{(1)}(z-y;m)
\nonumber \\
&&=-{i \over 6 } \gamma^a R^{(0) \mu}{}_{\!\rho\sigma 
a}(x-y)^\rho(x-y)^\sigma
 \partial_\mu\tilde{S}^{(0)}(x-y;m)
\nonumber \\
&&-\frac{i}{4}\gamma^a\sigma^{bc}R^{(0)}_{bca\lambda}(x-y)^\lambda
\tilde{S}^{(0)}(x-y;m),
\label{G0:spn2}
\end{eqnarray}
or, finally,
\begin{eqnarray}
&&\tilde{S}^{(0)}(x-y;m)=\int d^{D}z S_{00}(x-z;m)
\nonumber \\
&&\times\biggl[ -{i \over 6}
\gamma^a R^{(0) \mu}{}_{\!\rho\sigma a}(z-y)^\rho 
(z-y)^\sigma\partial_\mu
\tilde{S}^{(0)}(z-y;m)
\nonumber \\
&&-{i \over 4}\gamma^a \sigma^{bc}R^{(0)}_{bca\lambda}(z-y)^\lambda\biggr]
\tilde{S}_0(z-y;m).
\label{G0:spn3}
\end{eqnarray}

However, we need only the coincidence limit $x \to y$ to calculate
the effective potential. It provides us the opportunity to
simplify (\ref{G0:spn3}) especially for the constant curvature 
spacetimes (\ref{Rconst}).
Thus the expression for the GF, $\tilde{S}^{(1)}$,
in the spacetime 
with an arbitrary dimension $D$ is the following:
\begin{eqnarray}
\tilde{S}^{(1)}(0;m)&=&-\frac{iR}{12D(D-1)}\int d^{D}z S_{00}(-z; m)
\biggl[ 2\!\not{\!z}z^\mu\partial_\mu \tilde{S}^{(0)}(z;m)
\nonumber \\
&&-2z^2\gamma^\mu \partial_\mu \tilde{S}^{(0)}(z;m)
+3(D-1)\!\not{\!z}\tilde{S}^{(0)}(z;m)\biggr].
\label{Eqn:G1:spn}
\end{eqnarray}

Now we begin the calculation of the Green function for the $D=4$
Gross-Neveu model in the external magnetic field. 
According to the Schwinger proper-time method
as in the previous subsection
the flat spacetime GF is found to be \cite{K1,Sch}:
\begin{eqnarray}
&&\tilde{S}^{(0)}(z;m)=-i\int_0^\infty \frac{ds}{16(\pi s)^2}
\exp\left[-i\left(\frac{\pi}{2}+sm^2\right)\right] 
\exp\left(-\frac{i}{4s}z_\mu C^{\mu\nu}z_\nu\right)
\nonumber \\
&&\times\left(m+\frac{1}{2s}\gamma^\mu 
C_{\mu\nu}z^{\nu}-\frac{e}{2}\gamma^\mu
F_{\mu\nu}z^\nu\right)\biggl[ (eBs) \cot 
(eBs)-\frac{es}{2}\gamma^\mu\gamma^\nu
F_{\mu\nu} \biggr] ,
\label{4D:G1:spn}
\end{eqnarray}
where $C_{\mu\nu}$ is defined in Eq.(\ref{def:Cmn}).
For $S_{00}(-z;m)$ we have directly from (\ref{4D:G1:spn}):
\begin{equation}
S_{00}(-z;m)=-i \int_0^\infty{dt \over 16(\pi 
t)^2}\exp\left[-i\left({\pi\over 2}+
m^2 t+ \frac{z^2}{4t}\right)\right]
\biggl(m-\frac{\not{\!z}}{2t}\biggr).
\label{4D:GF}
\end{equation}

Substituting (\ref{4D:GF}) in (\ref{Eqn:G1:spn}) 
and calculating the trace over the spinor 
indices, we have :
\begin{eqnarray}
Sp\tilde{S}^{(1)}(0;m) &=& {-iRm\over 36}\int{d^{4}zdtds\over (16 
\pi^2 ts)^2}
\nonumber \\
&&\times\exp \left\{-i\biggl[ (t+s)m^2+{t+s\over 4ts}z_\parallel^2
-z^2_\bot{1+ eBt \cot (eBs) \over 4t}\biggr] \right\}
\nonumber \\
&&\times
\biggl\{eBs \cot (eBs)\biggl[ z_\parallel^2
\left(-{9\over 2t}+{7-4eBs \cot (eBs)\over 2s}\right)
\nonumber \\
&&+z^2_\bot \left({9\over 2t}+{4-7eBs \cot (eBs)\over 2s}\right)
\nonumber \\
&&-{i\over 2s^2}z_\parallel^2 z^2_\bot
(1-eBs \cot (eBs))^2\biggr]
-e^2 B^2\biggl[ s\left(2z_\parallel^2+\frac{7}{2}z^2_\bot\right)
\nonumber \\
&&+{i\over 2}z_\parallel^2 z^2_\bot(eBs \cot (eBs) -1) \biggr] 
\biggr\},
\label{34}
\end{eqnarray}
where
\begin{equation}
z^2_\bot=z_1^2+z_2^2, z^2_\parallel= z^2_0 - z^2_3 .
\label{tau}
\end{equation}
After the Wick rotation and integration over $z$, one gets:
\begin{eqnarray}
Sp\tilde{S}^{(1)}(0;m)& =& {iRm \over 96\pi^2} \int \frac{dtds}
{(t+s)^2 (1+ eBt\coth (eBs))^2}\exp[-(t+s)m^2]
\nonumber \\
&&\times\biggl[ eBt( eBt+eBs)+2( eBt+3eBs)\coth (eBs)
\nonumber \\
&&+2 eBt(eBs- eBt)\coth^2 (eBs)\biggr].
\label{4D:tildG}
\end{eqnarray}

The same program can be done for the 3D case. The GF in the flat 
space-time with the external constant magnetic field has 
the following form \cite{K1,K2}:
\begin{eqnarray}
\tilde{S}^{(0)}(z;m)&=&-i\int_0^\infty \frac{ds}{8(\pi s)^{3/2}}
 e^{-i({\pi /4}+sm^2)} exp\left(-\frac{i}{4s}z_\mu 
C^{\mu\nu}z_\nu\right)
\nonumber \\
&&\times\left(m+\frac{1}{2s}\gamma^\mu 
C_{\mu\nu}z^{nu}-\frac{e}{2}\gamma^\mu
F_{\mu\nu}z^\nu\right)
\nonumber \\
&&\times\biggl[ eBs \cot 
(eBs)-\frac{es}{2}\gamma^\mu\gamma^\nu
F_{\mu\nu} \biggr].
\label{3D:G0}
\end{eqnarray}
For $S_{00}(-z;m)$ we have directly from (\ref{3D:G0}):
\begin{equation}
S_{00}(-z;m)=-i \int_0^\infty{dt \over 8(\pi t)^{3/2}}
\exp\left[-i\left({\pi\over 4}+m^2 t+ \frac{z^2}{4t}\right)\right]
\biggl(m-\frac{\not{\!z}}{2t}\biggr).
\label{3D:G00}
\end{equation}
Substituting (\ref{3D:G0}),(\ref{3D:G00}) in (\ref{Eqn:G1:spn}), 
we have:
\begin{eqnarray}
Sp\tilde{S}^{(1)}(0;m)&=&{Rm\over 1152\pi^3
}\int{d^{3}zdtds\over (ts)^{3/ 2}}
\exp \biggl\{-i\biggl[ (t+s)m^2+{t+s\over 4ts}z_0^2
\nonumber \\
&&-z^2_\bot{1+ eBt \cot (eBs) \over 4t}\biggr] \biggr\}
\nonumber \\
&&\times\biggl\{eBs \cot (eBs)
\biggl[ z_0^2(-{3\over t}+{3-2eBs \cot (eBs)\over s})
\nonumber \\
&&+z^2_\bot ({3\over t}+{1-2eBs \cot (eBs)\over s})
\nonumber \\
&&-{i\over 2s^2}z_0^2 z^2_\bot
(1-eBs \cot (eBs))^2\biggr]
\nonumber \\
&&+e^2 B^2\biggl[-2s(z_0^2+z^2_\bot)
\nonumber \\
&&+{i\over 2}z_0^2 
z^2_\bot
(1-eBs \cot (eBs)) \biggr] \biggr\}.
\label{3DG1}
\end{eqnarray}
After the Wick rotation and integration over $z$, one gets:
\begin{eqnarray}
&&Sp\tilde{S}^{(1)}(0;m)={iRm \over 72\pi^{3/2}} \int 
{dtds \over(t+s)^{3/2}(1+ eBt\coth (eBs))^2}
\exp[-(t+s)m^2]\nonumber \\&&
\times\biggl[2 eBt( eBt+eBs)+(9eBs+5 eBt)
\coth (eBs)\nonumber \\&&
+ eBt(eBs-3 eBt)\coth^2 (eBs)\biggr].
\label{3DGF}
\end{eqnarray}

The proper-time integrations remain in our final expressions
of the scalar and spinor Green functions in an external magnetic 
and gravitational field. All remained integrands are 
exponentially suppressed at the limit $s \to \infty$
and/or $t \to \infty$. There are divergences at
$s \to 0$ and/or $t \to 0$ for $D \geq 2$.

\section{NJL model}

Let us discuss now the NJL model \cite{NJL}
in an external magnetic 
and gravitational field (for a review, see \cite{rev}).
The NJL model is one of the simplest models where dynamical 
symmetry breaking is possible. It is well-know that 
the chiral symmetry  of the model is dynamically broken 
down for a sufficiently large coupling constant. 
In the present paper we want to know the effect of an external 
electromagnetic and gravitational field to the dynamical
symmetry breaking.

The NJL model is defined by the action which is given by
\begin{equation}
S=\int d^D x \sqrt{-g} \biggl\{i \overline{\psi}\gamma^\mu (x)D_\mu \psi +
{\lambda \over 2N} \biggl[ (\overline{\psi}\psi)^2+
(\overline{\psi} i \gamma_5 \psi)^2 \biggr] \biggr\},
\label{Act:GN}
\end{equation}
where $N$ is the number of the fermions.
This action has the chiral $U(1)$ symmetry.
Introducing the auxiliary fields
\begin{equation}
\sigma=-{\lambda \over N }(\overline{\psi} \psi ), \pi=-{\lambda\over 
N }
\overline{\psi} i \gamma_5 \psi ,
\label{Aux:GN}
\end{equation}
we can rewrite the action (\ref{Act:GN}) as:
\begin{equation}
S=\int d^D x \sqrt{-g} \left[ i\overline{\psi}\gamma^\mu D_\mu \psi -
{N \over 2\lambda 
}(\sigma^2+\pi^2)-\overline{\psi}(\sigma+i\pi\gamma_5)\psi\right] .
\label{Act:Aux}
\end{equation}
In the leading order of the $1/N$ expansion
the effective action is given by
\begin{equation}
{1 \over N } \Gamma_{eff}(\sigma,\pi)=
-\int d^D x \sqrt{-g}\, {\sigma^2+\pi^2 \over 2\lambda} 
-i\ln \det [i\gamma^\mu (x)D_\mu-
(\sigma+i\gamma_5\pi)].
\label{V:Aux}
\end{equation}
We can put now $\pi=0$ because the final expression would depend on 
the combination $\sigma^2+ \pi^2$ only. In this case the effective
action of the NJL model has the same form as the one of the 
Gross-Neveu model \cite{GN}.
Note that Eq.(\ref{V:Aux}) represents the particular example of 
the effective action for composite fields \cite{CJT}.

Defining the effective potential as
$V_{eff} = -\Gamma_{eff}/ N \int d^D x\sqrt{-g} $ for constant 
configurations of $\sigma$ and $\pi$, we obtain
\begin{equation}
V_{eff}(\sigma)={\sigma^2 \over 2\lambda }+i Sp \ln
\langle x| [ i\gamma^\mu (x)D_\mu -\sigma] |x \rangle.
\label{v:njl}
\end{equation}
The second term of the right hand side in Eq.(\ref{v:njl})
is rewritten as
\begin{equation}
i Sp \ln
\langle x| [ i\gamma^\mu (x)D_\mu -\sigma] |x \rangle
=-i Sp \int^{\sigma} dm\ S(x,x;m).
\label{V:GF}
\end{equation}
Thus we can express the effective potential of the NJL model
by the spinor Green function $S(x,x,\sigma)$.
Substituting the Green function evaluated in the previous
section to Eq.(\ref{V:GF}) we obtain the effective potential 
in an external magnetic and gravitational field 

\begin{figure}[t]
    \begin{minipage}{.49\linewidth}
    \begin{center}
\setlength{\unitlength}{0.240900pt}
\begin{picture}(900,809)(0,0)
\thicklines \path(220,254)(240,254)
\thicklines \path(836,254)(816,254)
\put(198,254){\makebox(0,0)[r]{-0.015}}
\thicklines \path(220,410)(240,410)
\thicklines \path(836,410)(816,410)
\put(198,410){\makebox(0,0)[r]{-0.01}}
\thicklines \path(220,567)(240,567)
\thicklines \path(836,567)(816,567)
\put(198,567){\makebox(0,0)[r]{-0.005}}
\thicklines \path(220,723)(240,723)
\thicklines \path(836,723)(816,723)
\put(198,723){\makebox(0,0)[r]{0}}
\thicklines \path(220,113)(220,133)
\thicklines \path(220,786)(220,766)
\put(220,68){\makebox(0,0){0}}
\thicklines \path(357,113)(357,133)
\thicklines \path(357,786)(357,766)
\put(357,68){\makebox(0,0){0.2}}
\thicklines \path(494,113)(494,133)
\thicklines \path(494,786)(494,766)
\put(494,68){\makebox(0,0){0.4}}
\thicklines \path(631,113)(631,133)
\thicklines \path(631,786)(631,766)
\put(631,68){\makebox(0,0){0.6}}
\thicklines \path(768,113)(768,133)
\thicklines \path(768,786)(768,766)
\put(768,68){\makebox(0,0){0.8}}
\thicklines \path(220,113)(836,113)(836,786)(220,786)(220,113)
\put(45,899)
{\makebox(0,0)[l]{\shortstack{$V^{(0)}_{eff}(\sigma)/\mu^4$}}}
\put(528,23){\makebox(0,0){$\sigma/\mu$}}
\put(425,736){\makebox(0,0)[l]{{\small $eB=0$}}}
\put(357,191){\makebox(0,0)[l]{{\small $eB=2\mu^2$}}}
\thinlines \path(220,723)(220,723)(221,723)(227,723)(234,723)(247,723)
(273,720)(299,717)(325,712)(351,707)(377,701)(403,695)(429,689)
(455,684)(481,679)(494,678)(507,677)(514,677)(517,677)(519,677)
(520,677)(521,677)(522,677)(523,677)(523,677)(525,677)(527,677)
(533,677)(546,678)(559,679)(585,685)(611,695)(637,710)(663,730)
(689,756)(713,786)
\thinlines \path(220,723)(220,723)(221,723)(247,721)(273,715)(299,708)
(325,698)(351,688)(377,677)(403,666)(429,655)(455,645)(481,636)
(507,629)(533,624)(540,623)(546,623)(549,623)(553,623)(554,623)
(556,623)(557,623)(558,623)(558,623)(559,623)(561,623)(562,623)
(562,623)(566,623)(572,623)(585,624)(611,631)(637,641)(663,658)
(689,680)(715,710)(742,747)(764,786)
\thinlines \path(220,723)(220,723)(221,723)(253,717)(286,702)(318,683)
(351,660)(383,635)(416,610)(449,585)(481,562)(514,541)(546,525)
(579,514)(595,511)(603,510)(607,510)(609,510)(611,510)(612,510)
(613,510)(614,510)(615,510)(618,510)(620,510)(628,510)(644,513)
(677,526)(709,549)(742,584)(774,633)(807,697)(836,769)
\thinlines \path(220,723)(220,723)(221,723)(253,713)(286,692)(318,662)
(351,628)(383,590)(416,551)(449,512)(481,474)(514,438)(546,407)
(579,380)(611,360)(628,353)(644,349)(652,347)(656,346)(660,346)
(662,346)(664,346)(665,346)(666,346)(667,346)(668,346)(670,346)
(672,346)(677,346)(693,349)(709,354)(742,375)(774,410)(807,460)
(836,520)
\thinlines \path(220,723)(220,723)(221,723)(247,714)(273,694)(299,666)
(325,632)(351,594)(377,553)(403,509)(429,465)(455,421)(481,377)
(507,334)(533,293)(559,256)(585,221)(611,191)(637,167)(663,148)
(689,135)(702,132)(709,131)(712,131)(714,131)(715,130)(716,130)
(717,130)(718,130)(719,130)(720,130)(722,130)(728,131)(742,134)
(768,146)(802,177)(836,227)
\end{picture}
    (a) $V^{(0)}_{eff}$ for $R =0$.
    \end{center}
    \end{minipage}
\hfill
    \begin{minipage}{.49\linewidth}
    \begin{center}
\setlength{\unitlength}{0.240900pt}
\begin{picture}(900,809)(0,0)
\thicklines \path(220,174)(240,174)
\thicklines \path(836,174)(816,174)
\put(198,174){\makebox(0,0)[r]{0}}
\thicklines \path(220,297)(240,297)
\thicklines \path(836,297)(816,297)
\put(198,297){\makebox(0,0)[r]{0.001}}
\thicklines \path(220,419)(240,419)
\thicklines \path(836,419)(816,419)
\put(198,419){\makebox(0,0)[r]{0.002}}
\thicklines \path(220,541)(240,541)
\thicklines \path(836,541)(816,541)
\put(198,541){\makebox(0,0)[r]{0.003}}
\thicklines \path(220,664)(240,664)
\thicklines \path(836,664)(816,664)
\put(198,664){\makebox(0,0)[r]{0.004}}
\thicklines \path(220,786)(240,786)
\thicklines \path(836,786)(816,786)
\put(198,786){\makebox(0,0)[r]{0.005}}
\thicklines \path(220,113)(220,133)
\thicklines \path(220,786)(220,766)
\put(220,68){\makebox(0,0){0}}
\thicklines \path(357,113)(357,133)
\thicklines \path(357,786)(357,766)
\put(357,68){\makebox(0,0){0.2}}
\thicklines \path(494,113)(494,133)
\thicklines \path(494,786)(494,766)
\put(494,68){\makebox(0,0){0.4}}
\thicklines \path(631,113)(631,133)
\thicklines \path(631,786)(631,766)
\put(631,68){\makebox(0,0){0.6}}
\thicklines \path(768,113)(768,133)
\thicklines \path(768,786)(768,766)
\put(768,68){\makebox(0,0){0.8}}
\thicklines \path(220,113)(836,113)(836,786)(220,786)(220,113)
\put(45,899)
{\makebox(0,0)[l]{\shortstack{$V^{(1)}_{eff}(\sigma)/\mu^4$}}}
\put(528,23){\makebox(0,0){$\sigma/\mu$}}
\put(466,223){\makebox(0,0)[l]{{\small $eB=0$}}}
\put(302,443){\makebox(0,0)[l]{{\small $eB=2\mu^2$}}}
\thinlines \path(220,174)(220,174)(221,174)(242,176)(263,179)(284,183)
(305,189)(326,196)(348,204)(369,213)(390,223)(411,233)(432,245)
(453,257)(475,269)(496,282)(517,296)(538,310)(559,325)(580,340)
(601,356)(623,372)(644,388)(665,405)(689,425)(719,449)(748,473)
(777,499)(807,524)(836,550)
\thinlines \path(220,174)(220,174)(221,174)(245,180)(270,191)(294,203)
(318,216)(343,231)(367,246)(392,261)(416,278)(440,294)(465,312)
(489,329)(514,347)(538,366)(562,385)(587,404)(611,424)(636,444)
(660,464)(685,485)(709,505)(733,527)(777,566)(807,592)(836,619)
\thinlines \path(220,174)(220,174)(221,174)(250,183)(279,199)(309,218)
(338,239)(367,260)(396,283)(426,306)(455,331)(484,355)(514,380)
(543,405)(572,431)(602,457)(631,483)(660,510)(689,537)(719,564)
(748,591)(777,619)(807,647)(836,675)
\thinlines \path(220,174)(220,174)(221,174)(250,184)(279,202)(309,223)
(338,246)(367,270)(396,296)(426,322)(455,349)(484,377)(514,404)
(543,432)(572,460)(602,489)(631,518)(660,546)(689,575)(719,604)
(748,634)(777,663)(807,692)(836,722)
\thinlines \path(220,174)(220,174)(221,174)(250,185)(279,204)(309,227)
(338,252)(367,278)(396,306)(426,335)(455,364)(484,394)(514,424)
(543,454)(572,485)(602,515)(631,546)(660,577)(689,608)(719,639)
(748,670)(777,701)(807,732)(836,763)
\end{picture}
    (b) $V^{(1)}_{eff}$ for $R =0$.
    \end{center}
    \end{minipage}
\vglue 1ex
\caption{The behavior of $V^{(0)}_{eff}$ and $V^{(1)}_{eff}$
are shown with the varying magnetic field 
$eB (=0,\mu^2/2,\mu^2,3\mu^2/2,2\mu^2)$
for fixed $\lambda (=1/2.5)$
and fixed $\Lambda (=10\mu)$ 
in four dimensional flat spacetime.
$\mu$ is an arbitrary mass scale.
We normalize that $V_{eff}^{(0)}(0)=V_{eff}^{(1)}(0)=0$.}
\label{fig:v:njl1}
\end{figure}
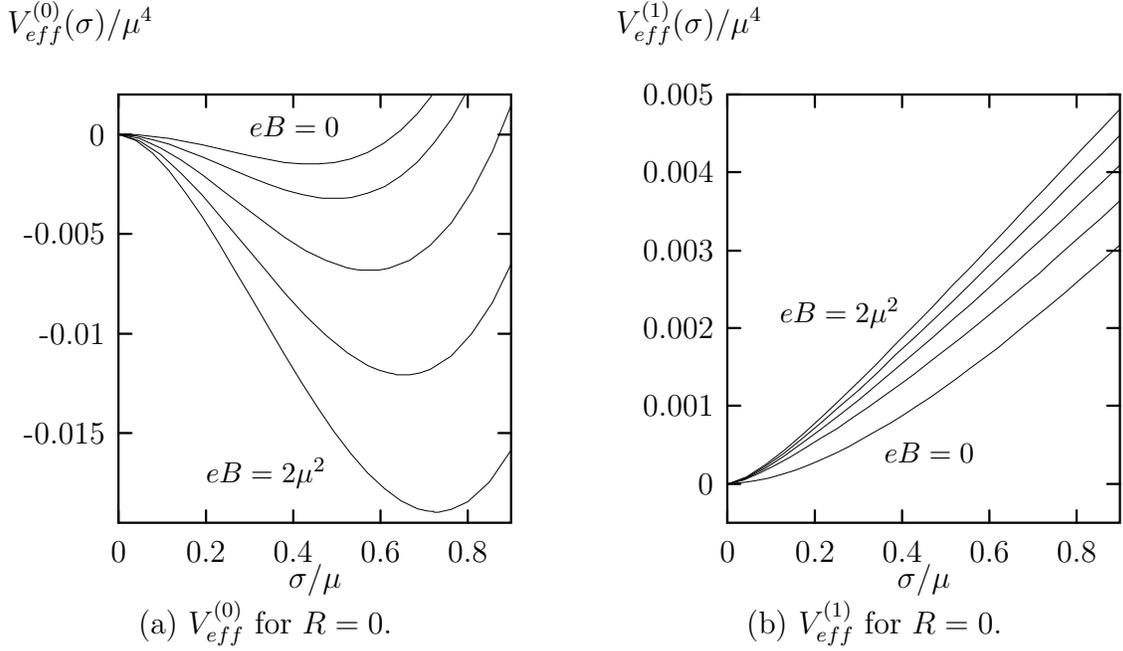

In four-dimensional spacetime the linear curvature correction
for the effective potential is given by
\begin{eqnarray}
&&V_{eff}^{(1)}(\sigma)=-{R\over 192\pi^2}\int_{1/\Lambda^2}^\infty
\int_{1/\Lambda^2}^\infty {dtds\over(t+s)^3(1+ eBt \coth (eBs) )^2}
\nonumber \\
&&\times\exp[-(t+s)\sigma^2]
[ eBt( eBt+eBs)
+2( eBt+3eBs)\coth (eBs)
\nonumber \\
&&+2 eBt(eBs- eBt)\coth^2 (eBs)],
\label{V:WCAPP}
\end{eqnarray}
where we introduce the proper-time cut-off $\Lambda$.
In four dimensions four-fermion models are not
renormalizable. Here we define the finite theory
by introducing the proper-time cut-off $\Lambda$.
Taking into account the proper-time representation for 
$V_{eff}$ in four-dimensional flat spacetime \cite{Sch}, 
we obtain the effective potential
with the linear curvature accuracy:
\begin{eqnarray}
&&V_{eff}(\sigma)={\sigma^2\over 2\lambda}+{1\over 8\pi^2}
\int_{1/\Lambda^2}^\infty {ds\over s^3} 
\exp(-s\sigma^2)eBs 
\coth (eBs)
\nonumber \\
&&-{R\over 192\pi^2}\int_{1/\Lambda^2}^\infty \int_{1/\Lambda^2}^\infty 
\frac{ds dt}{(t+s)^3(1+ eBt \coth (eBs))^2}
\nonumber \\
&&\times\exp 
[-(t+s)\sigma^2][ eBt( eBt+eBs)
+2( eBt+3eBs)\coth (eBs)
\nonumber \\
&&+ 2eBt(eBs- eBt)\coth^2 (eBs)].
\label{V:4D}
\end{eqnarray}
It should be emphasised here that for $B=0$ we obtain just the same 
expression for 
4D NJL model $V_{eff}$ in the proper-time 
representation as it has been found out in \cite{GGO}
(See also \cite{rev}). 

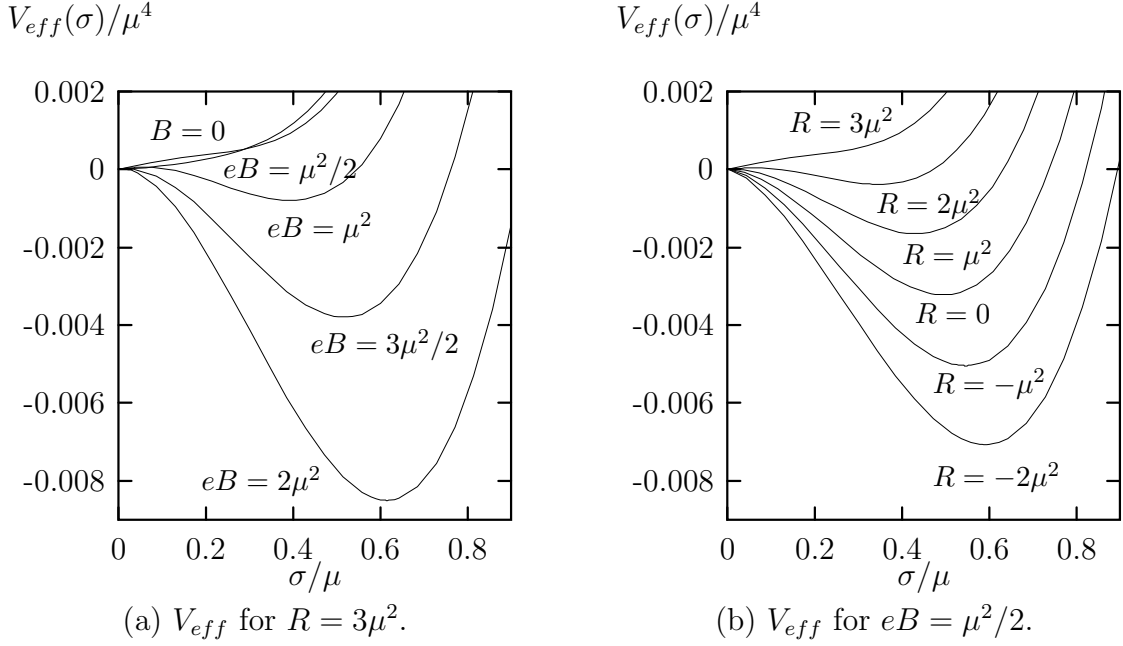
\begin{figure}
    \begin{minipage}{.49\linewidth}
    \begin{center}
\setlength{\unitlength}{0.240900pt}
\begin{picture}(900,809)(0,0)
\thicklines \path(220,174)(240,174)
\thicklines \path(836,174)(816,174)
\put(198,174){\makebox(0,0)[r]{-0.008}}
\thicklines \path(220,297)(240,297)
\thicklines \path(836,297)(816,297)
\put(198,297){\makebox(0,0)[r]{-0.006}}
\thicklines \path(220,419)(240,419)
\thicklines \path(836,419)(816,419)
\put(198,419){\makebox(0,0)[r]{-0.004}}
\thicklines \path(220,541)(240,541)
\thicklines \path(836,541)(816,541)
\put(198,541){\makebox(0,0)[r]{-0.002}}
\thicklines \path(220,664)(240,664)
\thicklines \path(836,664)(816,664)
\put(198,664){\makebox(0,0)[r]{0}}
\thicklines \path(220,786)(240,786)
\thicklines \path(836,786)(816,786)
\put(198,786){\makebox(0,0)[r]{0.002}}
\thicklines \path(220,113)(220,133)
\thicklines \path(220,786)(220,766)
\put(220,68){\makebox(0,0){0}}
\thicklines \path(357,113)(357,133)
\thicklines \path(357,786)(357,766)
\put(357,68){\makebox(0,0){0.2}}
\thicklines \path(494,113)(494,133)
\thicklines \path(494,786)(494,766)
\put(494,68){\makebox(0,0){0.4}}
\thicklines \path(631,113)(631,133)
\thicklines \path(631,786)(631,766)
\put(631,68){\makebox(0,0){0.6}}
\thicklines \path(768,113)(768,133)
\thicklines \path(768,786)(768,766)
\put(768,68){\makebox(0,0){0.8}}
\thicklines \path(220,113)(836,113)(836,786)(220,786)(220,113)
\put(45,899){\makebox(0,0)[l]{\shortstack{$V_{eff}(\sigma)/\mu^4$}}}
\put(528,23){\makebox(0,0){$\sigma/\mu$}}
\put(268,725){\makebox(0,0)[l]{{\small $B=0$}}}
\put(384,664){\makebox(0,0)[l]{{\small $eB=\mu^2/2$}}}
\put(453,572){\makebox(0,0)[l]{{\small $eB=\mu^2$}}}
\put(521,388){\makebox(0,0)[l]{{\small $eB=3\mu^2/2$}}}
\put(350,174){\makebox(0,0)[l]{{\small $eB=2\mu^2$}}}
\thinlines \path(220,664)(220,664)(221,664)(242,665)(263,666)(284,669)
(305,671)(326,674)(348,678)(369,682)(390,687)(411,693)(432,701)
(453,710)(475,722)(496,737)(517,755)(538,777)(545,786)
\thinlines \path(220,664)(220,664)(221,664)(245,669)(270,674)(294,678)
(318,682)(343,685)(367,688)(392,691)(416,695)(440,701)(465,709)
(489,720)(514,736)(538,756)(562,782)(565,786)
\thinlines \path(220,664)(220,664)(221,664)(235,666)(243,666)(246,667)
(248,667)(249,667)(250,667)(251,667)(252,667)(254,667)(257,667)
(259,667)(260,667)(261,667)(262,667)(263,667)(265,667)(268,667)
(272,667)(279,666)(309,662)(338,654)(367,645)(396,634)(426,625)
(455,618)(470,616)(477,615)(481,615)(484,615)(486,615)(487,615)
(488,615)(489,615)(490,615)(492,615)(494,615)(495,615)(499,615)
(514,617)(543,626)(572,643)(602,671)(631,712)(660,766)(668,786)
\thinlines \path(220,664)(220,664)(221,664)(228,664)(230,664)(232,665)
(233,665)(234,665)(234,665)(235,665)(236,665)(237,665)(239,664)
(243,664)(250,662)(279,653)(309,635)(338,613)(367,585)(396,556)
(426,526)(455,498)(484,472)(514,451)(543,437)(558,433)(565,432)
(569,432)(570,432)(572,432)(573,432)(574,432)(575,432)(576,432)
(578,432)(580,432)(587,433)(602,436)(631,453)(660,484)(689,532)
(719,597)(748,683)(776,786)
\thinlines \path(220,664)(220,664)(221,664)(224,664)(225,664)(226,664)
(227,664)(228,664)(229,664)(230,664)(232,664)(235,663)(239,663)
(243,661)(250,658)(279,637)(309,605)(338,565)(367,517)(396,465)
(426,411)(455,357)(484,305)(514,257)(543,215)(572,181)(602,156)
(616,149)(624,146)(631,144)(635,144)(636,144)(638,144)(639,144)
(640,143)(641,143)(642,143)(644,144)(646,144)(653,145)(660,147)
(689,165)(719,202)(748,259)(777,339)(807,444)(836,577)
\end{picture}
    (a) $V_{eff}$ for $R=3\mu^2$.
    \end{center}
    \end{minipage}
\hfill
    \begin{minipage}{.49\linewidth}
    \begin{center}
\setlength{\unitlength}{0.240900pt}
\begin{picture}(900,809)(0,0)
\thicklines \path(220,174)(240,174)
\thicklines \path(836,174)(816,174)
\put(198,174){\makebox(0,0)[r]{-0.008}}
\thicklines \path(220,297)(240,297)
\thicklines \path(836,297)(816,297)
\put(198,297){\makebox(0,0)[r]{-0.006}}
\thicklines \path(220,419)(240,419)
\thicklines \path(836,419)(816,419)
\put(198,419){\makebox(0,0)[r]{-0.004}}
\thicklines \path(220,541)(240,541)
\thicklines \path(836,541)(816,541)
\put(198,541){\makebox(0,0)[r]{-0.002}}
\thicklines \path(220,664)(240,664)
\thicklines \path(836,664)(816,664)
\put(198,664){\makebox(0,0)[r]{0}}
\thicklines \path(220,786)(240,786)
\thicklines \path(836,786)(816,786)
\put(198,786){\makebox(0,0)[r]{0.002}}
\thicklines \path(220,113)(220,133)
\thicklines \path(220,786)(220,766)
\put(220,68){\makebox(0,0){0}}
\thicklines \path(357,113)(357,133)
\thicklines \path(357,786)(357,766)
\put(357,68){\makebox(0,0){0.2}}
\thicklines \path(494,113)(494,133)
\thicklines \path(494,786)(494,766)
\put(494,68){\makebox(0,0){0.4}}
\thicklines \path(631,113)(631,133)
\thicklines \path(631,786)(631,766)
\put(631,68){\makebox(0,0){0.6}}
\thicklines \path(768,113)(768,133)
\thicklines \path(768,786)(768,766)
\put(768,68){\makebox(0,0){0.8}}
\thicklines \path(220,113)(836,113)(836,786)(220,786)(220,113)
\put(45,899){\makebox(0,0)[l]{\shortstack{$V_{eff}(\sigma)/\mu^4$}}}
\put(528,23){\makebox(0,0){$\sigma/\mu$}}
\put(316,737){\makebox(0,0)[l]{{\small $R=3\mu^2$}}}
\put(453,609){\makebox(0,0)[l]{{\small $R=2\mu^2$}}}
\put(494,529){\makebox(0,0)[l]{{\small $R=\mu^2$}}}
\put(514,437){\makebox(0,0)[l]{{\small $R=0$}}}
\put(542,327){\makebox(0,0)[l]{{\small $R=-\mu^2$}}}
\put(542,180){\makebox(0,0)[l]{{\small $R=-2\mu^2$}}}
\thinlines \path(220,664)(220,664)(221,664)(245,669)(270,674)(294,678)
(318,682)(343,685)(367,688)(392,691)(416,695)(440,701)(465,709)
(489,720)(514,736)(538,756)(562,782)(565,786)
\thinlines \path(220,664)(220,664)(221,664)(231,665)(242,665)(247,666)
(250,666)(252,666)(255,666)(256,666)(257,666)(258,666)(258,666)
(259,666)(260,666)(260,666)(263,666)(274,666)(284,665)(305,663)
(326,660)(348,656)(369,652)(390,648)(411,644)(432,641)(443,641)
(448,640)(451,640)(453,640)(455,640)(455,640)(456,640)(457,640)
(457,640)(459,640)(461,640)(464,640)(475,641)(496,644)(517,650)
(538,660)(559,674)(580,694)(601,718)(623,749)(644,786)
\thinlines \path(220,664)(220,664)(221,664)(224,664)(224,664)(225,664)
(226,664)(227,664)(228,664)(228,664)(230,664)(233,664)(239,663)
(245,663)(270,658)(294,650)(318,640)(343,629)(367,616)(392,604)
(416,592)(440,581)(465,572)(489,565)(501,563)(508,563)(511,563)
(512,563)(514,563)(514,563)(515,563)(516,563)(517,563)(517,563)
(518,563)(520,563)(526,563)(538,565)(562,572)(587,586)(611,607)
(636,636)(660,676)(685,726)(708,786)
\thinlines \path(220,664)(220,664)(221,664)(247,659)(273,648)(299,633)
(325,614)(351,594)(377,573)(403,551)(429,529)(455,510)(481,492)
(507,479)(533,470)(540,468)(546,467)(549,467)(553,467)(554,467)
(556,467)(557,467)(558,467)(558,467)(559,467)(561,467)(562,467)
(562,467)(566,467)(572,467)(585,470)(611,482)(637,503)(663,536)
(689,580)(715,638)(742,710)(764,786)
\thinlines \path(220,664)(220,664)(221,664)(250,654)(279,634)(309,608)
(338,578)(367,545)(396,511)(426,477)(455,444)(484,415)(514,389)
(543,370)(558,363)(572,358)(580,357)(583,356)(587,356)(589,356)
(591,356)(592,355)(592,355)(593,355)(594,355)(595,355)(596,355)
(598,356)(602,356)(616,359)(631,364)(660,386)(689,423)(719,476)
(748,549)(777,642)(807,760)(812,786)
\thinlines \path(220,664)(220,664)(221,664)(250,650)(279,623)(309,590)
(338,551)(367,509)(396,466)(426,422)(455,379)(484,339)(514,303)
(543,272)(572,249)(587,241)(602,235)(609,233)(616,232)(620,232)
(622,231)(623,231)(624,231)(624,231)(625,231)(627,231)(631,232)
(638,233)(646,235)(660,241)(689,265)(719,306)(748,366)(777,447)
(807,551)(836,680)
\end{picture}
    (b) $V_{eff}$ for $eB =\mu^{2}/2$.
    \end{center}
    \end{minipage}
\vglue 1ex
\caption{The behavior of the effective potential $V_{eff}$
is shown with the varying $B$ or $R$ for fixed $\lambda (=1/2.5)$
and fixed $\Lambda (=10\mu)$ in four dimensions.}
\label{fig:v:njl2}
\end{figure}

For $D=3$ the linear curvature correction for the effective potential
is given by
\begin{eqnarray}
&&V_{eff}^{(1)}(\sigma)=-{R\over 144\pi^{3/2}}\int_{1/\Lambda^2}^\infty
\int_{1/\Lambda^2}^\infty {dtds\over(t+s)^{5/2}
(1+ eBt \coth (eBs))^2}
\nonumber \\
&&\times\exp[-(t+s)\sigma^2]
[2 eBt( eBt+eBs)
+(9eBs+5 eBt)\coth (eBs)
\nonumber \\
&&+ eBt(eBs-3 eBt)\coth^2 (eBs)].
\label{3D:V}
\end{eqnarray}
For $D=3$, four-fermion models are known to be renormalizable 
in the sense of the $1/N$ expansion.
Thus the cut-off dependence in the effective potential
disappears after the usual renormalization procedure. 
Taking into account the proper-time representation for 
$V_{eff}$
in flat spacetime \cite{K1,K2}, 
we can write the effective potential with the 
linear curvature accuracy:
\begin{eqnarray}
&&V_{eff}(\sigma)={\sigma^2\over 2\lambda}+{1\over 4\pi^{3/2}}
\int_{1/\Lambda^2}^\infty {ds\over s^{5/2}}\exp(-s\sigma^2)eBs 
\coth (eBs)
\nonumber \\
&&-{R\over 144\pi^{3/2}}\int_{1/\Lambda^2}^\infty 
\int_{1/\Lambda^2}^\infty 
\frac{ds dt}{(t+s)^{5/2}(1+ eBt 
\coth (eBs))^2}
\nonumber \\
&&\times\exp[-(t+s)\sigma^2]\biggl[2 eBt( eBt+eBs)
+(9eBs+5 eBt)\coth (eBs)
\nonumber \\
&&+  eBt(eBs-3 eBt)\coth^2 (eBs)\biggr].
\label{V:3D:WC}
\end{eqnarray}
The effective potential (\ref{V:3D:WC}) gives again the correct 
expression for $V_{eff}$ of 3D 
NJL model 
in the proper-time representation with $B=0$ \cite{GOS}. 

Using Eq. (\ref{V:4D}) we numerically calculate the effective 
potential for $D=4$ and show it in Figs.1 and 2. In drawing
Fig.1 and 2 the coupling constant $\lambda$ is kept in the region
where the chiral symmetry is broken down for $R=B=0$. 
To see the curvature effect in the external constant magnetic 
field we divide the effective potential into the terms 
independent of the curvature and the terms linear in the 
curvature, $V_{eff}=V_{eff}^{(0)}+V_{eff}^{(1)}$.
In Fig.1 typical behaviors of the effective 
potential $V^{(0)}_{eff}$ and $V^{(1)}_{eff}$ are given for 
fixed four-fermion coupling
constant $\lambda$ and proper-time cut-off $\Lambda$.  
It is clearly seen in Figs.1 and 2.(a) that a constant magnetic 
field $B$ decreases the potential energy in the true vacuum 
even in the curved spacetime. Curvature effects depend
on the sign of the spacetime curvature $R$ as is shown
in Fig. 2.(b). A negative curvature decreases the potential energy. 
On the contrary, positive curvature increases the potential energy.
The broken chiral symmetry is restored for a sufficiently
large and positive curvature even in an external large magnetic 
field. 

\section{SUSY NJL model}

In the present section we consider the supersymmetric NJL 
model in the external constant magnetic field and 
external gravity. 

In flat space such a model has been introduced in Ref.\cite{SNJL}.
Generalization of the model for the case of non-minimal coupling
with the external gravitational field (via scalar-gravitational
coupling constants) has been presented in Ref.\cite{BIO}.
We consider the model of Ref.\cite{BIO} in curved spacetime with
the external magnetic field.
The action in components is given by
\begin{eqnarray}
     S&=&\displaystyle\int d^{4}x\sqrt{-g}\biggl[
         -\phi^{\dag}(D^{\mu}D_{\mu}
         +\sigma^{2}+\xi_{1} R)\phi
         -{\phi^{c}}^{\dag}(D^{\mu}D_{\mu}
         +\sigma^{2}+\xi_{2} R)\phi^{c}\nonumber \\
       &&\displaystyle +\bar{\psi}(i\gamma^{\mu}D_{\mu}
         -\sigma)\psi-\frac{1}{2\lambda}\sigma^{2}\biggr],
\label{s:snjl2}
\end{eqnarray}
where $\sigma$ is an auxiliary
scalar as in the original NJL model,
$\psi$ is $N$ component Dirac spinor.

Note that actually the action (\ref{s:snjl2}) represents
SUSY NJL model non-minimally interacting with the external 
gravity and minimally interacting with the external
magnetic field.
It is evident that supersymmetry of SUSY NJL model
is always broken in the external fields.
So it maybe natural to call the theory with action 
(\ref{s:snjl2}) the extended NJL model.

At the leading order of the $1/N$ expansion the effective 
potential of the SUSY NJL model (\ref{s:snjl2}) is given by
\begin{eqnarray}
     &&V_{eff}(\sigma)=\frac{1}{2\lambda}\sigma^{2}
     +i Sp \ln
     \langle x| [ i\gamma^\mu (x)D_\mu -\sigma] |x \rangle
     \nonumber \\
     &&-i Sp \ln
     \langle x| [ D^\mu D_\mu +\sigma^{2}+\xi_{1} R] |x \rangle
     \nonumber \\
     &&-i Sp \ln
     \langle x| [ D^\mu D_\mu +\sigma^{2}+\xi_{2} R] |x \rangle .
\label{v:susynjl}
\end{eqnarray}
The second term in the right hand side of Eq.(\ref{v:susynjl})
corresponds to the radiative correction by spinor fields and
is rewritten by the spinor GF as is evaluated in
the previous section (See Eq.(\ref{V:GF})).
The third and the fourth terms in the right hand side of 
Eq.(\ref{v:susynjl}) 
correspond to the radiative correction by scalar fields and
is represented by the scalar GF
\begin{eqnarray}
      - i Sp \ln
     \langle x| [ D^\mu D_\mu +\sigma^{2}+\xi R] |x \rangle
     \nonumber \\
     =2 i \int^{\sigma} m\ dm\ G(x,x,m).
\label{splnG}
\end{eqnarray}
Because of the supersymmetry both the radiative correction 
by spinor fields and the one by scalar fields
are cancelled out in flat spacetime with vanishing
electromagnetic fields. Thus the vacuum expectation value
of $\sigma$ disappears for $R=B=0$.
Substituting Eqs.(\ref{SpG0}) and (\ref{SpG1:FIN}) 
into Eq.(\ref{splnG}) and integrating over $m$ one gets
\begin{eqnarray}
     &&2 i \int^{\sigma} m\ dm\ G(x,x,m)
     \nonumber \\
     &&=-\frac{1}{(4\pi)^{D/2}}\int_{0}^{\infty}ds s^{-D/2}
     \frac{eB}{\sinh(eBs)}
     \exp \left(-\sigma^{2}s\right)
     \nonumber \\
     &&+\frac{1}{(4\pi)^{D/2}}\frac{R}{D(D-1)}
     \int_{0}^{\infty} \int_{0}^{\infty}dt\ ds 
     \exp \left[-(s+t)\sigma^{2}\right]
     \nonumber \\
     &&\times\frac{(s+t)^{-D/2}}
     {1+eBt\coth(eBs)}\frac{eB}{\sinh(eBs)}
     \nonumber \\
     &&\times \left\{\left[
     -\frac{D-3}{6}
     \left(D-2+2eBs\coth(eBs)\right)
     +\frac{-2D+1}{3}\right]\frac{(D-2)t}{s+t}\right.
     \nonumber \\
     &&+\left[
     -\frac{D-3}{6s}
     \left(D-2+2eBs\coth(eBs)\right)
     +\frac{-2D+1}{3}eB\coth(eBs)\right]
     \nonumber \\
     &&\times\frac{2t}{1+eBt\coth(eBs)}
     \nonumber \\
     &&+\frac{2(D-1)}{3}\left[\frac{D(D-2)}{4}
     \left(\frac{t}{s+t}\right)^{2}
     +2\left(\frac{eBt\coth(eBs)}{1+eBt\coth(eBs)}\right)^{2}\right]
     \nonumber \\
     &&+\frac{1}{3s}
     \left[(D-3)\left((eBs)^{2}\coth^{2}(eBs)+1\right)
     +4eBs\coth(eBs)\right]
     \nonumber \\
     &&\times\left.\frac{(D-2)t}{s+t}
     \frac{t}{1+eBt\coth(eBs)}+D(D-1)\xi\right\} .
\label{int:G:susy}
\end{eqnarray}
In some special dimensions Eq.(\ref{int:G:susy})
simplifies. Inserting the Eq.(\ref{int:G:susy}) into
Eq.(\ref{v:susynjl}) we obtain the final expression
of the effective potential of the SUSY NJL model.
Therefore the effective potential for $D=4$ reads
\begin{eqnarray}
     &&V_{eff}(\sigma)=\frac{1}{2\lambda}\rho^{2}
     +{1\over 8\pi^2}
     \int_{1/\Lambda^2}^\infty {ds\over s^3} \exp(-s\sigma^2)eBs 
     \coth (eBs)
     \nonumber \\
     &&-{R\over 192\pi^2}\int_{1/\Lambda^2}^\infty 
     \int_{1/\Lambda^2}^\infty 
     \frac{ds dt}{(s+t)^3(1+ eBt \coth (eBs))^2}
     \nonumber \\
     &&\times\exp [-(s+t)\sigma^2]
     [ eBt( eBt+eBs)
     \nonumber \\
     &&+2( eBt+3eBs)\coth (eBs)
     +2eBt(eBs- eBt)\coth^2 (eBs)]
     \nonumber \\
     &&-\frac{1}{8\pi^{2}}\int_{1/\Lambda^2}^{\infty}\frac{ds}{s^{3}}
     \frac{eBs}{\sinh(eBs)}
     \exp \left(-s\sigma^{2}\right)
     \nonumber \\
     &&+\frac{R}{48\pi^{2}}\int_{1/\Lambda^2}^\infty 
     \int_{1/\Lambda^2}^\infty 
     \frac{ds dt}{(s+t)^2(1+ eBt \coth (eBs))}
     \nonumber \\
     &&\times \frac{eB}{\sinh(eBs)}
     \exp [-(s+t)\sigma^{2}]
     \left\{\left[
     -\frac{4}{3}-\frac{1}{6}eBs\coth(eBs)
     \right]\frac{2t}{s+t}\right.
     \nonumber \\
     &&+\left[
     -\frac{1}{6s}-\frac{4}{3}eB\coth(eBs)\right]
     \frac{2t}{1+eBt\coth(eBs)}
     \nonumber \\
     &&+2\left[
     \left(\frac{t}{s+t}\right)^{2}
     +\left(\frac{eBt\coth(eBs)}{1+eBt\coth(eBs)}\right)^{2}\right]
     \nonumber \\
     &&+\frac{1}{3s}
     \left[(eBs)^{2}\coth^{2}(eBs)+1
     +4eBs\coth(eBs)\right]
     \nonumber \\
     &&\left. \times\frac{t}{s+t}
     \frac{t}{1+eBt\coth(eBs)}
     +3(\xi_{1}+\xi_{2})\right\} ,
\label{v:susynjl:4D}
\end{eqnarray}
where we regularize the effective potential by introducing
the proper-time cut-off $\Lambda$.
The effective potential (\ref{v:susynjl:4D}) reproduces
the expression for $V_{eff}(\sigma)$
in Ref.\cite{BIO} at the limit $B\rightarrow 0$.
The proper-time integrations in Eq.(\ref{v:susynjl:4D}) are
numerically performed below.

\begin{figure}
    \begin{minipage}{.49\linewidth}
    \begin{center}
\setlength{\unitlength}{0.240900pt}
\begin{picture}(900,809)(0,0)
\thicklines \path(220,113)(240,113)
\thicklines \path(836,113)(816,113)
\put(198,113){\makebox(0,0)[r]{-0.004}}
\thicklines \path(220,197)(240,197)
\thicklines \path(836,197)(816,197)
\put(198,197){\makebox(0,0)[r]{-0.003}}
\thicklines \path(220,281)(240,281)
\thicklines \path(836,281)(816,281)
\put(198,281){\makebox(0,0)[r]{-0.002}}
\thicklines \path(220,365)(240,365)
\thicklines \path(836,365)(816,365)
\put(198,365){\makebox(0,0)[r]{-0.001}}
\thicklines \path(220,450)(240,450)
\thicklines \path(836,450)(816,450)
\put(198,450){\makebox(0,0)[r]{0}}
\thicklines \path(220,534)(240,534)
\thicklines \path(836,534)(816,534)
\put(198,534){\makebox(0,0)[r]{0.001}}
\thicklines \path(220,618)(240,618)
\thicklines \path(836,618)(816,618)
\put(198,618){\makebox(0,0)[r]{0.002}}
\thicklines \path(220,702)(240,702)
\thicklines \path(836,702)(816,702)
\put(198,702){\makebox(0,0)[r]{0.003}}
\thicklines \path(220,786)(240,786)
\thicklines \path(836,786)(816,786)
\put(198,786){\makebox(0,0)[r]{0.004}}
\thicklines \path(220,113)(220,133)
\thicklines \path(220,786)(220,766)
\put(220,68){\makebox(0,0){0}}
\thicklines \path(374,113)(374,133)
\thicklines \path(374,786)(374,766)
\put(374,68){\makebox(0,0){0.05}}
\thicklines \path(528,113)(528,133)
\thicklines \path(528,786)(528,766)
\put(528,68){\makebox(0,0){0.1}}
\thicklines \path(682,113)(682,133)
\thicklines \path(682,786)(682,766)
\put(682,68){\makebox(0,0){0.15}}
\thicklines \path(836,113)(836,133)
\thicklines \path(836,786)(836,766)
\put(836,68){\makebox(0,0){0.2}}
\thicklines \path(220,113)(836,113)(836,786)(220,786)(220,113)
\put(45,899)
{\makebox(0,0)[l]{\shortstack{$V^{(0)}_{eff}(\sigma)/\mu^4$}}}
\put(528,23){\makebox(0,0){$\sigma/\mu$}}
\put(251,727){\makebox(0,0)[l]{{\small $eB$}}}
\put(257,677){\makebox(0,0)[l]{{\small $=0$}}}
\put(528,197){\makebox(0,0)[l]{{\small $eB=15\mu^2$}}}
\thinlines \path(220,450)(220,450)(223,450)(238,453)(253,462)(269,476)
(284,495)(299,519)(314,548)(329,582)(345,621)(360,666)(375,715)
(390,770)(394,786)
\thinlines \path(220,450)(220,450)(223,450)(232,450)(240,452)(258,458)
(275,467)(293,479)(310,497)(327,520)(345,547)(362,579)(380,617)
(397,659)(414,705)(432,758)(440,786)
\thinlines \path(220,450)(220,450)(223,450)(236,450)(242,450)(248,450)
(252,450)(253,450)(255,450)(255,450)(256,450)(257,450)(257,450)
(258,449)(259,448)(259,448)(259,447)(260,447)(261,447)(261,447)
(262,447)(262,447)(263,447)(263,447)(264,447)(266,447)(267,447)
(270,446)(272,446)(274,446)(274,446)(275,446)(275,446)(276,446)
(276,446)(277,446)(278,446)(278,446)(280,446)(286,447)(299,448)
(312,450)(324,453)(337,458)(349,463)(362,470)(375,479)(387,489)
(400,500)(413,513)(425,528)
\thinlines \path(425,528)(438,544)(451,562)(463,582)(476,604)(488,627)
(501,652)(514,679)(526,708)(539,740)(552,773)(556,786)
\thinlines \path(220,450)(220,450)(223,449)(267,432)(311,399)(354,362)
(398,327)(442,295)(486,271)(508,262)(519,259)(524,257)(530,256)
(535,255)(536,255)(538,255)(539,255)(540,255)(543,255)(546,255)
(551,254)(554,254)(557,254)(558,254)(560,254)(561,254)(562,254)
(564,254)(565,253)(568,253)(573,253)(595,257)(617,264)(661,289)
(705,330)(748,389)(792,466)(836,563)
\end{picture}
    (a) $V^{(0)}_{eff}$ for $R=0$.
    \end{center}
    \end{minipage}
\hfill
    \begin{minipage}{.49\linewidth}
    \begin{center}
\setlength{\unitlength}{0.240900pt}
\begin{picture}(900,809)(0,0)
\thicklines \path(220,165)(240,165)
\thicklines \path(836,165)(816,165)
\put(198,165){\makebox(0,0)[r]{0}}
\thicklines \path(220,268)(240,268)
\thicklines \path(836,268)(816,268)
\put(198,268){\makebox(0,0)[r]{0.002}}
\thicklines \path(220,372)(240,372)
\thicklines \path(836,372)(816,372)
\put(198,372){\makebox(0,0)[r]{0.004}}
\thicklines \path(220,475)(240,475)
\thicklines \path(836,475)(816,475)
\put(198,475){\makebox(0,0)[r]{0.006}}
\thicklines \path(220,579)(240,579)
\thicklines \path(836,579)(816,579)
\put(198,579){\makebox(0,0)[r]{0.008}}
\thicklines \path(220,682)(240,682)
\thicklines \path(836,682)(816,682)
\put(198,682){\makebox(0,0)[r]{0.01}}
\thicklines \path(220,786)(240,786)
\thicklines \path(836,786)(816,786)
\put(198,786){\makebox(0,0)[r]{0.012}}
\thicklines \path(220,113)(220,133)
\thicklines \path(220,786)(220,766)
\put(220,68){\makebox(0,0){0}}
\thicklines \path(357,113)(357,133)
\thicklines \path(357,786)(357,766)
\put(357,68){\makebox(0,0){0.2}}
\thicklines \path(494,113)(494,133)
\thicklines \path(494,786)(494,766)
\put(494,68){\makebox(0,0){0.4}}
\thicklines \path(631,113)(631,133)
\thicklines \path(631,786)(631,766)
\put(631,68){\makebox(0,0){0.6}}
\thicklines \path(768,113)(768,133)
\thicklines \path(768,786)(768,766)
\put(768,68){\makebox(0,0){0.8}}
\thicklines \path(220,113)(836,113)(836,786)(220,786)(220,113)
\put(45,899)
{\makebox(0,0)[l]{\shortstack{$V^{(1)}_{eff}(\sigma)/\mu^4$}}}
\put(528,23){\makebox(0,0){$\sigma/\mu$}}
\put(466,217){\makebox(0,0)[l]{{\small $eB=0$}}}
\put(357,486){\makebox(0,0)[l]{{\small $eB=15\mu^2$}}}
\thinlines \path(220,165)(220,165)(221,165)(242,167)(263,171)(284,177)
(305,185)(326,194)(348,204)(369,216)(390,229)(411,244)(432,259)
(453,275)(475,292)(496,309)(517,328)(538,347)(559,367)(580,388)
(601,409)(623,431)(644,453)(665,476)(699,514)(733,553)(768,593)
(802,634)(836,676)
\thinlines \path(220,165)(220,165)(221,165)(245,169)(270,178)(294,190)
(318,203)(343,219)(367,236)(392,254)(416,273)(440,293)(465,315)
(489,337)(514,360)(538,384)(562,408)(587,434)(611,459)(636,486)
(660,513)(685,541)(709,569)(733,598)(768,639)(802,681)(836,723)
\thinlines \path(220,165)(220,165)(221,165)(250,171)(279,184)(309,201)
(338,220)(367,242)(396,266)(426,291)(455,318)(484,346)(514,375)
(543,405)(572,436)(602,468)(631,501)(660,534)(689,568)(719,603)
(748,639)(777,675)(807,711)(836,748)
\thinlines \path(220,165)(220,165)(221,165)(250,172)(279,185)(309,203)
(338,223)(367,246)(396,271)(426,297)(455,325)(484,354)(514,384)
(543,415)(572,447)(602,480)(631,513)(660,548)(689,582)(719,618)
(748,654)(777,690)(807,727)(836,764)
\end{picture}
    (b) $V^{(1)}_{eff}$
        for $R=0$ and $\xi_{1}+\xi_{2} =0$.
    \end{center}
    \end{minipage}
\vglue 10ex
    \begin{minipage}{.49\linewidth}
    \begin{center}
\setlength{\unitlength}{0.240900pt}
\begin{picture}(900,809)(0,0)
\thicklines \path(220,113)(240,113)
\thicklines \path(836,113)(816,113)
\put(198,113){\makebox(0,0)[r]{-0.008}}
\thicklines \path(220,225)(240,225)
\thicklines \path(836,225)(816,225)
\put(198,225){\makebox(0,0)[r]{-0.006}}
\thicklines \path(220,337)(240,337)
\thicklines \path(836,337)(816,337)
\put(198,337){\makebox(0,0)[r]{-0.004}}
\thicklines \path(220,450)(240,450)
\thicklines \path(836,450)(816,450)
\put(198,450){\makebox(0,0)[r]{-0.002}}
\thicklines \path(220,562)(240,562)
\thicklines \path(836,562)(816,562)
\put(198,562){\makebox(0,0)[r]{0}}
\thicklines \path(220,674)(240,674)
\thicklines \path(836,674)(816,674)
\put(198,674){\makebox(0,0)[r]{0.002}}
\thicklines \path(220,786)(240,786)
\thicklines \path(836,786)(816,786)
\put(198,786){\makebox(0,0)[r]{0.004}}
\thicklines \path(220,113)(220,133)
\thicklines \path(220,786)(220,766)
\put(220,68){\makebox(0,0){0}}
\thicklines \path(357,113)(357,133)
\thicklines \path(357,786)(357,766)
\put(357,68){\makebox(0,0){0.2}}
\thicklines \path(494,113)(494,133)
\thicklines \path(494,786)(494,766)
\put(494,68){\makebox(0,0){0.4}}
\thicklines \path(631,113)(631,133)
\thicklines \path(631,786)(631,766)
\put(631,68){\makebox(0,0){0.6}}
\thicklines \path(768,113)(768,133)
\thicklines \path(768,786)(768,766)
\put(768,68){\makebox(0,0){0.8}}
\thicklines \path(220,113)(836,113)(836,786)(220,786)(220,113)
\put(45,899)
{\makebox(0,0)[l]{\shortstack{$V^{(1)}_{eff}(\sigma)/\mu^4$}}}
\put(528,23){\makebox(0,0){$\sigma/\mu$}}
\put(398,281){\makebox(0,0)[l]{{\small $eB=0$}}}
\put(288,730){\makebox(0,0)[l]{{\small $eB=15\mu^2$}}}
\thinlines \path(220,562)(220,562)(221,562)(242,560)(263,556)(284,550)
(305,542)(326,533)(348,522)(369,510)(390,498)(411,484)(432,469)
(453,454)(475,438)(496,421)(517,403)(538,385)(559,366)(580,347)
(601,327)(623,307)(644,286)(665,265)(699,230)(733,194)(768,157)
(802,120)(808,113)
\thinlines \path(220,562)(220,562)(221,562)(245,565)(270,571)(294,577)
(318,584)(343,589)(367,594)(392,598)(416,601)(440,603)(453,604)
(459,604)(465,604)(468,604)(469,604)(471,604)(472,604)(472,604)
(473,604)(474,604)(475,604)(477,604)(483,604)(489,604)(514,603)
(538,600)(562,597)(587,592)(611,587)(636,580)(660,573)(685,564)
(709,555)(733,545)(768,529)(802,511)(836,492)
\thinlines \path(220,562)(220,562)(221,562)(250,567)(279,577)(309,589)
(338,600)(367,612)(396,623)(426,633)(455,642)(484,651)(514,658)
(543,664)(572,669)(602,673)(631,675)(646,676)(660,676)(668,677)
(671,677)(673,677)(675,677)(676,677)(677,677)(678,677)(679,677)
(680,677)(682,677)(689,677)(704,676)(719,675)(748,673)(777,670)
(807,665)(836,660)
\thinlines \path(220,562)(220,562)(221,562)(250,568)(279,579)(309,593)
(338,607)(367,622)(396,637)(426,651)(455,665)(484,678)(514,691)
(543,702)(572,713)(602,723)(631,732)(660,739)(689,746)(719,752)
(748,757)(777,760)(807,763)(836,765)
\end{picture}
    (c) $V^{(1)}_{eff}$
        for $R=0$ and $\xi_{1}+\xi_{2} =1$.
    \end{center}
    \end{minipage}
\hfill
\vglue 1ex
\caption{The behavior of the effective potential $V^{(0)}_{eff}$
and $V^{(1)}_{eff}$
are shown with the varying magnetic field 
$eB (=0,5\mu^2,10\mu^2,15\mu^2)$
for fixed $\lambda (=1/2.5)$, 
and fixed $\Lambda (=10\mu)$ in four dimensions.
$\mu$ is an arbitrary mass scale.
We normalize that $V_{eff}(0)=0$.}
\label{fig:v:susynjl}
\end{figure}

\begin{figure}
    \begin{minipage}{.49\linewidth}
    \begin{center}
\setlength{\unitlength}{0.240900pt}
\begin{picture}(900,809)(0,0)
\thicklines \path(220,203)(240,203)
\thicklines \path(836,203)(816,203)
\put(198,203){\makebox(0,0)[r]{0}}
\thicklines \path(220,315)(240,315)
\thicklines \path(836,315)(816,315)
\put(198,315){\makebox(0,0)[r]{0.005}}
\thicklines \path(220,427)(240,427)
\thicklines \path(836,427)(816,427)
\put(198,427){\makebox(0,0)[r]{0.01}}
\thicklines \path(220,539)(240,539)
\thicklines \path(836,539)(816,539)
\put(198,539){\makebox(0,0)[r]{0.015}}
\thicklines \path(220,651)(240,651)
\thicklines \path(836,651)(816,651)
\put(198,651){\makebox(0,0)[r]{0.02}}
\thicklines \path(220,764)(240,764)
\thicklines \path(836,764)(816,764)
\put(198,764){\makebox(0,0)[r]{0.025}}
\thicklines \path(220,113)(220,133)
\thicklines \path(220,786)(220,766)
\put(220,68){\makebox(0,0){0}}
\thicklines \path(374,113)(374,133)
\thicklines \path(374,786)(374,766)
\put(374,68){\makebox(0,0){0.05}}
\thicklines \path(528,113)(528,133)
\thicklines \path(528,786)(528,766)
\put(528,68){\makebox(0,0){0.1}}
\thicklines \path(682,113)(682,133)
\thicklines \path(682,786)(682,766)
\put(682,68){\makebox(0,0){0.15}}
\thicklines \path(836,113)(836,133)
\thicklines \path(836,786)(836,766)
\put(836,68){\makebox(0,0){0.2}}
\thicklines \path(220,113)(836,113)(836,786)(220,786)(220,113)
\put(45,899){\makebox(0,0)[l]{\shortstack{$V_{eff}(\sigma)/\mu^4$}}}
\put(528,23){\makebox(0,0){$\sigma/\mu$}}
\put(420,651){\makebox(0,0)[l]{{\small $eB=0$}}}
\put(559,180){\makebox(0,0)[l]{{\small $eB=15\mu^2$}}}
\thinlines \path(220,203)(220,203)(220,203)(250,206)(279,215)(308,229)
(338,249)(367,274)(396,305)(426,342)(455,384)(484,431)(513,484)
(543,542)(572,606)(601,675)(631,750)(644,786)
\thinlines \path(220,203)(220,203)(220,203)(250,205)(279,211)(308,222)
(338,236)(367,255)(396,278)(426,306)(455,337)(484,373)(513,414)
(543,459)(572,508)(601,562)(631,621)(660,684)(689,751)(704,786)
\thinlines \path(220,203)(220,203)(220,203)(250,204)(279,206)(308,210)
(338,217)(367,226)(396,237)(426,252)(455,269)(484,290)(513,313)
(543,339)(572,369)(601,402)(631,438)(660,478)(689,522)(719,568)
(748,618)(777,672)(807,729)(834,786)
\thinlines \path(220,203)(220,203)(220,203)(250,203)(279,200)(308,197)
(338,195)(367,194)(396,194)(426,194)(455,196)(484,199)(513,204)
(543,210)(572,218)(601,229)(631,241)(660,255)(689,273)(719,291)
(748,312)(777,336)(807,362)(836,391)
\end{picture}
    (a) $V_{eff}$ for $R=5\mu^2$ and $\xi_{1}+\xi_{2} =0$.
    \end{center}
    \end{minipage}
\hfill
    \begin{minipage}{.49\linewidth}
    \begin{center}
\setlength{\unitlength}{0.240900pt}
\begin{picture}(900,809)(0,0)
\thicklines \path(220,203)(240,203)
\thicklines \path(836,203)(816,203)
\put(198,203){\makebox(0,0)[r]{0}}
\thicklines \path(220,315)(240,315)
\thicklines \path(836,315)(816,315)
\put(198,315){\makebox(0,0)[r]{0.005}}
\thicklines \path(220,427)(240,427)
\thicklines \path(836,427)(816,427)
\put(198,427){\makebox(0,0)[r]{0.01}}
\thicklines \path(220,539)(240,539)
\thicklines \path(836,539)(816,539)
\put(198,539){\makebox(0,0)[r]{0.015}}
\thicklines \path(220,651)(240,651)
\thicklines \path(836,651)(816,651)
\put(198,651){\makebox(0,0)[r]{0.02}}
\thicklines \path(220,764)(240,764)
\thicklines \path(836,764)(816,764)
\put(198,764){\makebox(0,0)[r]{0.025}}
\thicklines \path(220,113)(220,133)
\thicklines \path(220,786)(220,766)
\put(220,68){\makebox(0,0){0}}
\thicklines \path(374,113)(374,133)
\thicklines \path(374,786)(374,766)
\put(374,68){\makebox(0,0){0.05}}
\thicklines \path(528,113)(528,133)
\thicklines \path(528,786)(528,766)
\put(528,68){\makebox(0,0){0.1}}
\thicklines \path(682,113)(682,133)
\thicklines \path(682,786)(682,766)
\put(682,68){\makebox(0,0){0.15}}
\thicklines \path(836,113)(836,133)
\thicklines \path(836,786)(836,766)
\put(836,68){\makebox(0,0){0.2}}
\thicklines \path(220,113)(836,113)(836,786)(220,786)(220,113)
\put(45,899){\makebox(0,0)[l]{\shortstack{$V_{eff}(\sigma)/\mu^4$}}}
\put(528,23){\makebox(0,0){$\sigma/\mu$}}
\put(451,651){\makebox(0,0)[l]{{\small $eB=0$}}}
\put(559,169){\makebox(0,0)[l]{{\small $eB=15\mu^2$}}}
\thinlines \path(220,203)(220,203)(220,203)(250,205)(279,212)(308,223)
(338,239)(367,259)(396,284)(426,314)(455,349)(484,388)(513,432)
(543,481)(572,535)(601,593)(631,657)(660,725)(685,786)
\thinlines \path(220,203)(220,203)(220,203)(250,205)(279,211)(308,220)
(338,233)(367,251)(396,272)(426,297)(455,326)(484,359)(513,396)
(543,438)(572,483)(601,533)(631,586)(660,644)(689,706)(719,773)
(724,786)
\thinlines \path(220,203)(220,203)(220,203)(250,204)(279,205)(308,209)
(338,214)(367,222)(396,233)(426,245)(455,261)(484,279)(513,299)
(543,323)(572,350)(601,379)(631,412)(660,448)(689,487)(719,529)
(748,575)(777,624)(807,676)(836,732)
\thinlines \path(220,203)(220,203)(220,203)(250,203)(279,199)(308,196)
(338,194)(367,191)(396,190)(426,189)(455,189)(484,190)(513,193)
(543,197)(572,203)(601,210)(631,219)(660,231)(689,244)(719,259)
(748,277)(777,296)(807,319)(836,343)
\end{picture}
    (b) $V_{eff}$ for $R=5\mu^2$ and $\xi_{1}+\xi_{2} =1$.
    \end{center}
    \end{minipage}
\vglue 10ex
    \begin{minipage}{.49\linewidth}
    \begin{center}
\setlength{\unitlength}{0.240900pt}
\begin{picture}(900,809)(0,0)
\thicklines \path(220,161)(240,161)
\thicklines \path(836,161)(816,161)
\put(198,161){\makebox(0,0)[r]{-0.006}}
\thicklines \path(220,257)(240,257)
\thicklines \path(836,257)(816,257)
\put(198,257){\makebox(0,0)[r]{-0.004}}
\thicklines \path(220,353)(240,353)
\thicklines \path(836,353)(816,353)
\put(198,353){\makebox(0,0)[r]{-0.002}}
\thicklines \path(220,450)(240,450)
\thicklines \path(836,450)(816,450)
\put(198,450){\makebox(0,0)[r]{0}}
\thicklines \path(220,546)(240,546)
\thicklines \path(836,546)(816,546)
\put(198,546){\makebox(0,0)[r]{0.002}}
\thicklines \path(220,642)(240,642)
\thicklines \path(836,642)(816,642)
\put(198,642){\makebox(0,0)[r]{0.004}}
\thicklines \path(220,738)(240,738)
\thicklines \path(836,738)(816,738)
\put(198,738){\makebox(0,0)[r]{0.006}}
\thicklines \path(220,113)(220,133)
\thicklines \path(220,786)(220,766)
\put(220,68){\makebox(0,0){0}}
\thicklines \path(374,113)(374,133)
\thicklines \path(374,786)(374,766)
\put(374,68){\makebox(0,0){0.05}}
\thicklines \path(528,113)(528,133)
\thicklines \path(528,786)(528,766)
\put(528,68){\makebox(0,0){0.1}}
\thicklines \path(682,113)(682,133)
\thicklines \path(682,786)(682,766)
\put(682,68){\makebox(0,0){0.15}}
\thicklines \path(836,113)(836,133)
\thicklines \path(836,786)(836,766)
\put(836,68){\makebox(0,0){0.2}}
\thicklines \path(220,113)(836,113)(836,786)(220,786)(220,113)
\put(45,899){\makebox(0,0)[l]{\shortstack{$V_{eff}(\sigma)/\mu^4$}}}
\put(528,23){\makebox(0,0){$\sigma/\mu$}}
\put(343,642){\makebox(0,0)[l]{{\small $R=10\mu^2$}}}
\put(466,401){\makebox(0,0)[l]{{\small $R=5\mu^2$}}}
\put(528,286){\makebox(0,0)[l]{{\small $R=0$}}}
\put(337,185){\makebox(0,0)[l]{{\small $R=-5\mu^2$}}}
\thinlines \path(220,450)(220,450)(220,450)(250,452)(279,451)(308,454)
(338,460)(367,468)(396,479)(426,494)(455,513)(484,535)(513,562)
(543,594)(572,629)(601,670)(631,716)(660,767)(670,786)
\thinlines \path(220,450)(220,450)(220,450)(250,449)(279,443)(308,438)
(338,434)(367,431)(396,430)(426,431)(455,435)(484,441)(513,452)
(543,466)(572,483)(601,505)(631,531)(660,562)(689,599)(719,638)
(748,684)(777,735)(804,786)
\thinlines \path(220,450)(220,450)(220,449)(250,447)(279,434)(308,422)
(338,408)(367,394)(396,380)(426,368)(455,357)(484,348)(513,341)
(543,338)(572,337)(601,340)(631,347)(660,357)(689,373)(719,391)
(748,415)(777,443)(807,476)(836,514)
\thinlines \path(220,450)(220,450)(220,449)(250,444)(279,426)(308,405)
(338,382)(367,357)(396,331)(426,304)(455,279)(484,254)(513,231)
(543,211)(572,191)(601,175)(631,162)(660,153)(689,148)(719,144)
(748,145)(777,151)(807,161)(836,176)
\end{picture}
    (c) $V_{eff}$ for $eB=15\mu^2$ and $\xi_{1}+\xi_{2} =0$.
    \end{center}
    \end{minipage}
\hfill
    \begin{minipage}{.49\linewidth}
    \begin{center}
\setlength{\unitlength}{0.240900pt}
\begin{picture}(900,809)(0,0)
\thicklines \path(220,161)(240,161)
\thicklines \path(836,161)(816,161)
\put(198,161){\makebox(0,0)[r]{-0.006}}
\thicklines \path(220,257)(240,257)
\thicklines \path(836,257)(816,257)
\put(198,257){\makebox(0,0)[r]{-0.004}}
\thicklines \path(220,353)(240,353)
\thicklines \path(836,353)(816,353)
\put(198,353){\makebox(0,0)[r]{-0.002}}
\thicklines \path(220,450)(240,450)
\thicklines \path(836,450)(816,450)
\put(198,450){\makebox(0,0)[r]{0}}
\thicklines \path(220,546)(240,546)
\thicklines \path(836,546)(816,546)
\put(198,546){\makebox(0,0)[r]{0.002}}
\thicklines \path(220,642)(240,642)
\thicklines \path(836,642)(816,642)
\put(198,642){\makebox(0,0)[r]{0.004}}
\thicklines \path(220,738)(240,738)
\thicklines \path(836,738)(816,738)
\put(198,738){\makebox(0,0)[r]{0.006}}
\thicklines \path(220,113)(220,133)
\thicklines \path(220,786)(220,766)
\put(220,68){\makebox(0,0){0}}
\thicklines \path(374,113)(374,133)
\thicklines \path(374,786)(374,766)
\put(374,68){\makebox(0,0){0.05}}
\thicklines \path(528,113)(528,133)
\thicklines \path(528,786)(528,766)
\put(528,68){\makebox(0,0){0.1}}
\thicklines \path(682,113)(682,133)
\thicklines \path(682,786)(682,766)
\put(682,68){\makebox(0,0){0.15}}
\thicklines \path(836,113)(836,133)
\thicklines \path(836,786)(836,766)
\put(836,68){\makebox(0,0){0.2}}
\thicklines \path(220,113)(836,113)(836,786)(220,786)(220,113)
\put(45,899){\makebox(0,0)[l]{\shortstack{$V_{eff}(\sigma)/\mu^4$}}}
\put(528,23){\makebox(0,0){$\sigma/\mu$}}
\put(405,642){\makebox(0,0)[l]{{\small $R=10\mu^2$}}}
\put(466,392){\makebox(0,0)[l]{{\small $R=5\mu^2$}}}
\put(528,305){\makebox(0,0)[l]{{\small $R=0$}}}
\put(359,185){\makebox(0,0)[l]{{\small $R=-5\mu^2$}}}
\thinlines \path(220,450)(220,450)(220,449)(250,452)(279,449)(308,450)
(338,452)(367,456)(396,462)(426,471)(455,483)(484,497)(513,515)
(543,537)(572,561)(601,591)(631,624)(660,661)(689,705)(719,750)
(739,786)
\thinlines \path(220,450)(220,450)(220,449)(250,449)(279,442)(308,436)
(338,430)(367,425)(396,421)(426,419)(455,420)(484,422)(513,428)
(543,438)(572,450)(601,465)(631,485)(660,509)(689,539)(719,571)
(748,608)(777,650)(807,698)(836,750)
\thinlines \path(220,450)(220,450)(220,449)(250,447)(279,434)(308,422)
(338,408)(367,394)(396,380)(426,368)(455,357)(484,348)(513,341)
(543,338)(572,337)(601,340)(631,347)(660,357)(689,373)(719,391)
(748,415)(777,443)(807,476)(836,514)
\thinlines \path(220,450)(220,450)(220,450)(250,444)(279,427)(308,408)
(338,386)(367,363)(396,339)(426,316)(455,294)(484,273)(513,255)
(543,239)(572,225)(601,215)(631,208)(660,205)(689,208)(719,211)
(748,221)(777,235)(807,255)(836,279)
\end{picture}
    (d) $V_{eff}$ for $eB=15\mu^2$ and $\xi_{1}+\xi_{2} =1$.
    \end{center}
    \end{minipage}
\vglue 1ex
\caption{The behavior of the effective potential $V_{eff}$
are shown with the varying $B$ or $R$ for fixed $\lambda (=1/2.5)$, 
and fixed $\Lambda (=10\mu)$ in four dimensions.}
\label{fig:v:susynjl2}
\end{figure}
For $D=3$ the effective potential reads
\begin{eqnarray}
     &&V_{eff}(\sigma)={\sigma^2\over 2\lambda}+{1\over 4\pi^{3/2}}
     \int_{1/\Lambda^2}^\infty {ds\over s^{5/2}}\exp(-s\sigma^2)
     eBs \coth (eBs)
     \nonumber \\
     &&-{R\over 144\pi^{3/2}}\int_{1/\Lambda^2}^\infty 
     \int_{1/\Lambda^2}^\infty 
     \frac{ds dt}{(t+s)^{5/2}(1+ eBt 
     \coth (eBs))^2}
     \nonumber \\
     &&\times\exp[-(t+s)\sigma^2]
     \biggl[2 eBt( eBt+eBs)
     \nonumber \\
     &&+(9eBs+5 eBt)\coth (eBs)
     + eBt(eBs-3 eBt)\coth^2 (eBs)\biggr]
     \nonumber \\
     &&-{1\over 4\pi^{3/2}}
     \int_{1/\Lambda^2}^\infty {ds\over s^{5/2}}
     \exp(-s\sigma^2)\frac{eBs}{\sinh(eBs)}
     \nonumber \\
     &&+\frac{R}{24\pi^{3/2}}\int_{1/\Lambda^2}^\infty 
     \int_{1/\Lambda^2}^\infty 
     \frac{ds dt}{(t+s)^{3/2}(1+ eBt \coth (eBs))}
     \nonumber \\
     &&\times \frac{eB}{\sinh(eBs)}
       \exp [-(s+t)\sigma^{2}]
     \left\{-\frac{5}{3}
     \frac{t}{s+t}-\frac{5}{3}
     \frac{2eBt\coth(eBs)}{1+eBt\coth(eBs)}\right.
     \nonumber \\
     &&+\left(\frac{t}{s+t}\right)^{2}
     +\frac{8}{3}\left(\frac{eBt\coth(eBs)}
     {1+eBt\coth(eBs)}\right)^{2}
     \nonumber \\
     &&\left. +\frac{4}{3}\frac{t}{s+t}
     \frac{eBt\coth(eBs)}{1+eBt\coth(eBs)}
     +3(\xi_{1}+\xi_{2})\right\} .
\label{v:susynjl:3D}
\end{eqnarray}

Since the contribution of scalar fields and fermion fields 
are canceled in a supersymmetric theory only a symmetric 
phase is realized for $R=B=0$. The external gravitational 
and magnetic fields considered here break the supersymmetry 
of the theory. Thus we expect that the chiral symmetry may be 
broken down. Using Eq.(\ref{v:susynjl:4D}) we numerically 
calculate the effective potential of the SUSY NJL model 
for $D=4$ to study the phase structure of the theory.

In Figs.3 and 4 we illustrate the typical behaviors of the effective 
potential for $D=4$ in the SUSY NJL model.
In Fig.3 we draw the terms independent of R, $V^{(0)}$, and
the linear curvature correction to the 
effective potential, $V^{(1)}$. 
The linear curvature correction strongly
depends on the coupling
constant $\xi_{1}+\xi_{2}$.
There are large cancellations 
between the corrections of fermion and scalar fields. 
Thus the terms of the linear curvature correction 
in the SUSY NJL model are significally smaller than that 
in the NJL model. 
As is clearly seen in Fig.4 the chiral symmetry is broken
down for a sufficiently large magnetic field and/or a
negative curvature for $R=5\mu^2$ or $B=15\mu^{2}$.
The external curvature has the opposite effects in the
case of large $\xi_{1}+\xi_{2} (>1/2)$ for a small $B$
and/or large $\sigma$ (See Fig.3.(c)).

\section{Conclusion}

In the present paper we discussed the combined effect 
of the gravitational and magnetic fields to the chiral
symmetry breaking in NJL and SUSY NJL models. Chiral symmetry
is broken at non- zero and non- positive curvature. On the same time,
positive curvature acts against chiral symmetry breaking. 
Nevertheless, the magnetic field effects may be significally stronger
in realistic situations corresponding to early Universe with
primordial magnetic fields.

It should be noted that the effective potential calculated 
in the present paper depends on how to introduce cut-off 
parameter in the proper-time integral, though we do not develop 
it any further here.

Using the results presented in Ref. \cite{rev} it is not difficult to
calculate the effective potential in above two models exactly on the 
constant curvature spacetimes (DeSitter or anti- DeSitter, for 
example \cite{rev,ELOS}). Such calculation shows that curvature effects
estimated in this work are taken into account qualitatively correctly
(one can consider then very strong curvature as well). However
it is not clear how to make such calculation exactly both on curvature 
and on magnetic field. The only possibility to do so is to work on some
gravitational- magnetic background where exact solutions of field 
equations are known (for example, on conformal spacetime with 
magnetic field).

Another interesting proposal could be to study the dynamical symmetry 
breaking on electromagnetic- gravitational background representing
a combination of constant electromagnetic field with gravitational 
wave (it is known the effective Lagrangian in pure electromagnetic
or gravitational wave is trivial and no effect is expected). 
Such background may be roughly considered as signal which comes 
from strongly- graviting object with strong magnetic field. 
Then one may speculate on the possible use of the dynamical symmetry
breaking in gravitational waves detectors.

\vspace{8mm}

We thank I.~L.~Buchbinder, S.~J.~Gates and T.~Muta for helpful 
discussions.
T.~I. was supported in part by Monbusho Grant-in-Aid
for Scientific Research Fellowship, No.2616.
S.~D.~O. thanks COLCIENCIES (Colombis) and JSPS (Japan) for
partial support of this work.
This work (Yu.~I. ~Sh. and S.~D.~O.)  was supported in part by 
Ministerio de Educacion y Cultura (Spain). Yu.~I.~Sh.  also expresses his
deep gratitude to A.~Letwin and R.~Patov for their kind support.


\begin{thebibliography}{99}
\bibitem{NJL}  Y.~Nambu and G.~Jona-Lasinio,
               {\it Phys. Rev.} {\bf 122}, (1961) 345.          
\bibitem{GN}   D.~Gross and A.~Neveu, 
               {\it Phys. Rev.} {\bf D10}, (1974) 3235.
\bibitem{IMO}  T.~Inagaki, T.~Muta and S.~D.~Odintsov,
               {\it Mod. Phys. Lett.}, {\bf A8}, (1993) 2117.
\bibitem{EOS1} E.~Elizalde, S.~D.~Odintsov and Yu~.I.~Shil'nov,
               {\it Mod. Phys. Lett.} {\bf A9}, (1994) 913.
\bibitem{rev}  T.~Inagaki, T.~Muta and S.~D.~Odintsov, 
               {\it Dynamical Symmetry Breaking
                     in Curved Spacetime},
               to be published in 
               {\it Prog. Theor. Phys. (Supplment)}.
\bibitem{IN}   T.~Inagaki,
               {\it Int. J. Mod. Phys.} {\bf A11} (1996), 4561. 
\bibitem{ELOS} T.~Inagaki, S.Mukaigawa and T.Muta,
               {\it Phys. Rev.} {\bf D52}, (1996) R4267;\\
               E.~Elizalde, S.~Leseduarte, S.~D.~Odintsov and Yu.~I.~Shil'nov,
               {\it Phys. Rev.} {\bf D53}, (1996) 1917;\\
               K.~Ishikawa, T.~Inagaki and T.Muta,
               {\it Mod. Phy. Lett} {\bf A11}, (1996) 939.\\
               T.~Inagaki and K.~Ishikawa,
               {Thermal and Curvature Effects to the
               Dynamical Symmetry Breaking},
               to be published in 
               {\it Phys. Rev.} {\bf D}.
               G.~Miele and P.~Vitale, {\it Nucl. Phys.}, {\bf B494}
               (1997), 365 
\bibitem{HS}   T.~Muta and S.~D.~Odintsov,
               {\it Mod. Phys. Lett.} {\bf A6}, (1991) 3641;\\
               C.~T.~Hill and D.~S.~Salopek,
               {\it Ann. Phys. (NY).} {\bf 213}, (1992) 21.
\bibitem{M1}   M.~S.Turner and L.~M.~Widrow,
               {\it Phys. Rev.} {\bf D37}, (1988) 2743;\\
               T.~Vachaspati,
               {\it Phys. Lett.} {\bf B263}, (1991) 258;\\
               K.~Enqvist and P.~Olesen,
               {\it Phys. Lett.} {\bf B319}, (1993) 178;\\
               M.~Gasperini, M.~Giovannini and
               G.~Veneziano,
               {\it Phys. Rev. Lett.} {\bf 75}, (1995) 3796;\\
               G. Baum, D. Bodeker and L.~McLerran,
               {\it Phys. Rev.} {\bf D53}, (1996) 662;\\
               J.~M.~Cornwall, in {\it Unified Symmetry,}
                ( ed. by B.~N.~Kursunoglu et. al.,
                Plenum, N.Y., 1995.)
\bibitem{M2}   S.~P.~Klevansky and R.~H.~Lemmer,
               {\it Phys. Rev.} {\bf D38}, (1988) 3559;\\
               H.~Suganuma and T.~Tatsumi,
               {\it Ann. of Phys. (N.Y.)} {\bf 208}, (1991) 470;
               {\it Progr. Theor. Phys.} {\bf 90}, (1993) 379;\\
               I.~Krive and S.~Naftulin,
               {\it Phys. Rev.} {\bf D46}, (1992) 2337;\\
               S.~P.~Klevansky,
               {\it Rev. Mod. Phys.} {\bf 64}, (1992) 1;\\
               D.~Cangemi, E.~D.'Hoker and G.~V.~Dunne,
               {\it Phys. Rev.} {\bf D51}, (1995) 2513.
\bibitem{K1}   K.~G.~Klimenko, {\it Theor. Math. Phys. (in Russinan)}
               {\bf 89} (1991), 211;\\
               {\it Z. Phys.} {\bf C54}, (1992) 323;\\
               A.~S.~Vshivtsev, B.~V.~Magnitski and K.~G.~Klimenko,
               {\it JETP Lett.} {\bf 62} (1995) 283. 
\bibitem{K2}   C.~N.~Leung, Y.~J.~Ng and A.~W.~Ackly, {\it Phys. Rev.} 
               {\bf D54}, (1996), 4181;\\
               D.-~S.~Lee, C.~N.~Leung, Y.~J.~Ng, hep- th/9701172;\\
               M.~Ishi- i, T.~Kashiwa and N.~Tanemura, hep- th/9707248;\\
               S.~Kanemura, H.~Sato and H.~Tochimura, hep- th/9707285
\bibitem{GOS}  D.~M.~Gitman, S.~D.~Odintsov and Yu.~I.~Shil'nov, 
               {\it Phys. Rev.} {\bf D54}, (1996) 2968.
\bibitem{GGO}  B.~Geyer, L.~N.~Granda and S.~D.~Odintsov, 
               {\it Mod. Phys. Lett.} {\bf A11}, (1996) 2053.
               I.~Brevik, D.~M.~Gitman, S.~D.~Odintsov, hep-th/9611138;
               {\it Gravitation and Cosmology}, to appear 
\bibitem{BP}   T.~S.~Bunch and L.~Parker, 
               {\it Phys. Rev.} {\bf D20}, (1979) 2449.
\bibitem{Ein}  A.~Z.~Petrov, {\it Einstein Spaces}
               (Pergamon, Oxford, 1969).
\bibitem{PT}   L.~Parker and D.~Toms, 
               {\it Phys. Rev.} {\bf D29}, (1984) 1584.
\bibitem{Sch}  J.~Schwinger, 
               {\it Phys. Rev.} {\bf 82}, (1951) 664.
\bibitem{IZ}   C.~Itzykson and J.-B.~Zuber,
               ``Quantum Field Theory'', (McGraw-Hill Inc., 1980).
\bibitem{BG}   V.~G.~Bagrov and D.~M.~Gitman,
               {\it Exact Solution of Relativistic Wave
               Equation}, (Kluwer Ac. Pub. Dorolrecht, 1990.)
\bibitem{CJT}  J.~M.~Cornwall, R.~Jackiw and E.~Tomboulis, 
               {\it Phys. Rev.}, {\bf D10}, (1974) 2428.
\bibitem{SNJL} W.~Buchmuller and S.~T.~Love,
               {\it Nucl. Phys.} {\bf B204}, (1982) 213.
\bibitem{BIO}  I.~L.~Buchbinder, T.~Inagaki and S.~D.~Odintsov,
               hep-th/9702097.
\bibitem{BOS} I.~L.~Buchbinder, S.~D.~Odintsov, I.~L.~Shapiro,
              {\it Effective Action in Quantum Gravity}, 
              (IOP Publishing, Bristol and Philadelphia, 1992.) 
\end{thebibliography}
\end{document}